\documentclass[twocolumn]{aastex63}

\usepackage{graphicx}	
\usepackage{amsmath}	
\usepackage{amssymb}	
\usepackage{multirow}
\usepackage{tikz}
\usetikzlibrary{calc,trees,positioning,arrows,chains,shapes.geometric,%
    decorations.pathreplacing,decorations.pathmorphing,shapes,%
    matrix,shapes.symbols}

\received{June 1, 2019}
\revised{January 10, 2019}
\accepted{\today}
\submitjournal{AJ}

\shorttitle{A Multivariate Damped Random Walk Process}
\shortauthors{Hu and Tak}

\begin{document}

\title{Modeling Stochastic Variability in Multi-Band Time Series Data}

\correspondingauthor{Hyungsuk Tak}\email{tak@psu.edu}

\author{Zhirui Hu}
\affiliation{Department of Statistics, Harvard University, Cambridge, MA 02138, USA}

\author[0000-0003-0334-8742]{Hyungsuk Tak}
\affiliation{Center for Astrostatistics,  Pennsylvania State University, University Park, PA 16802, USA}
\affiliation{Department of Statistics,  Pennsylvania State University, University Park, PA 16802, USA}
\affiliation{Department of Astronomy and Astrophysics,  Pennsylvania State University, University Park, PA 16802, USA}
\affiliation{Institute for Computational and Data Sciences,  Pennsylvania State University, University Park, PA 16802, USA}



\begin{abstract}

In preparation for the era of the time-domain astronomy with upcoming large-scale surveys, we propose a state-space representation of a multivariate damped random walk process as a tool to analyze irregularly-spaced multi-filter light curves with heteroscedastic measurement errors. We adopt a computationally efficient and scalable Kalman-filtering approach to evaluate the likelihood function, leading to maximum $O(k^3n)$ complexity, where $k$ is the  number of available bands  and $n$ is the number of unique observation times across the $k$ bands. This is a significant computational advantage over a commonly used univariate Gaussian process that can stack up all multi-band light curves in one vector with maximum $O(k^3n^3)$ complexity. Using such efficient likelihood computation, we provide both maximum likelihood estimates and Bayesian posterior samples of the model parameters. Three numerical illustrations are presented: (i) analyzing simulated five-band light curves for a comparison with independent single-band fits; (ii) analyzing  five-band light curves of a quasar obtained from the Sloan Digital Sky Survey (SDSS) Stripe~82 to estimate  short-term variability and timescale; (iii) analyzing gravitationally lensed $g$- and $r$-band light curves of Q0957+561 to infer the time delay. Two R packages, \texttt{Rdrw} and \texttt{timedelay}, are publicly available to fit the proposed models. 

\end{abstract}



\section{Introduction} \label{sec1}

A Gaussian process (GP) is one of the most important data analytic tools in astronomy due to its well-known computational and mathematical conveniences. GPs are especially useful  for analyzing astronomical time series data \textcolor{black}{since} they are continuous-time processes accounting for irregular observation cadences. Moreover, GP's state-space representation  enables modeling heteroscedastic measurement errors as well. Such analytic advantages have made GPs so popular that it is nearly impossible to list all sub-fields of astronomy where GPs are useful; 19,483 ApJ articles appear on the webpage of IOPscience and 18,038 MNRAS articles show up on the webpage of MNRAS  with the keyword ``Gaussian process'' (on Feb 19, 2020).  However, it is the case that a \textcolor{black}{\emph{multi-output}} GP, which is \textcolor{black}{suitable} for modeling multi-band time series data, has not been well documented in the astronomical literature. \textcolor{black}{This vector-output GP is widely used in other fields, e.g., cokriging or coregionalization  in geostatistics \citep{journel1978, gelfand2004, alvarez2012kernels} and multi-task learning in machine learning \citep{caruana1997}. The key idea is to model dependence among multi-source data via a  covariance function to  take advantage of their dependent structure in making an inference or a prediction.}

We propose a state-space representation of a  multivariate damped random walk process as a specific class of a \textcolor{black}{multi-output} GP. This process is also called a multivariate Ornstein-Uhlenbeck process \citep{gardiner2009, singh2018fast} and a vectorized continuous-time auto-regressive model with order one, i.e., a vectorized {CAR(1)} or {CARMA(1, 0)} \citep{marquardt2007mCARMA}. In particular, the proposed is a multivariate generalization of the work of \cite{kelly2009variations}. They adopt a univariate GP with the Mat\'ern(1/2) covariance function (i.e., damped random walk process) to fit single-filter quasar light curves. Using \textcolor{black}{this} single-band model, they investigate associations between model parameters and  physical properties of quasars. Following their work,  \cite{macleod2010modeling}, \cite{kozlowski2010quantifying},  \cite{kim2012}, and \cite{andrae2013} show more empirical evidence for such astrophysical interpretations on the model parameters. The proposed multivariate generalization of their analytic tools can incorporate more data from all available bands into one comprehensive model. This enables more accurate inference on such physically meaningful model parameters.





\textcolor{black}{Also,} the multivariate aspect of the proposed is essential for studying stochastic AGN variability  in the  era of Vera C.~Rubin Observatory Legacy Survey of Space and Time  \citep[LSST,][]{ivezi2019lsst}.  LSST lightcurves are supposed to be  sparse when only one band is considered. In general, it is challenging to extract  information about short-term variability and timescale from sparsely observed single-band light curves. This problem becomes worse if the actual timescale of an AGN variability is much smaller than the typical observation cadence in each band. The proposed multi-band model, however, can alleviate this issue of sparse sampling. \textcolor{black}{This is} because it can take  advantage of  more data points  observed at non-overlapping times \textcolor{black}{in all} bands. It will  lead to more accurate inference on short-term variabilit\textcolor{black}{ies and timescales, which in turn} will be helpful for investigating AGN variability and light curve classification. 

From a methodological point of view, the proposed method is flexible enough to model various aspects of astronomical multi-filter light curves. Above all, the proposed process is a continuous-time process in a state-space representation, suitable for modeling irregularly-spaced multi-band time series data with heteroscedastic measurement errors.  Also, the process does not require the data at each observation time to be a vector of the same length; the number of observations at each observation time can range from one to $k$, the total number of bands.  This feature is desirable because there can be ties in observation times possibly due to rounding. In addition, the lengths of multi-filter light curves do not need to be the same; a light curve from one band can be longer than other light curves from  different bands.  Such a flexibility makes the proposed process ideal for modeling SDSS and LSST multi-band time series data with heteroscedastic measurement errors.

Moreover, the proposed process has a couple of computational advantages.  We adopt a Kalman-filtering approach for evaluating the resulting likelihood function with maximum $O(k^3n)$ complexity, where $n$ is the total number of unique observation times across the entire $k$ bands. This is  more scalable and efficient than an existing strategy of applying a univariate GP  to multi-filter light curves that are stacked up in a single vector \citep{zu2016app, czekala2017}. \textcolor{black}{This is because the univariate GP approach leads to maximum $O(k^3n^3)$ complexity}. Such an efficient likelihood computation makes the following likelihood-based inference efficient; we provide both maximum likelihood estimates and Bayesian posterior samples of  model parameters.

Our  numerical studies include a simulation study and two realistic data analyses. These show that the proposed process results in more comprehensive inference compared to independent single-band analyses. The simulation study generates multi-band light curves using the proposed  model with some fixed parameter values. Then it checks whether the resulting inference successfully recovers these generative parameter values. The next numerical illustration  analyzes realistic five-band light curves of a quasar obtained from SDSS Stripe 82 to infer short-term variabilities and timescales. Finally, we apply the proposed process to inferring the time delay between doubly lensed images of Q0957+561 whose light curves are observed in $g$- and $r$-band.

\section{Model specification}\label{modelspecification}
A univariate damped random walk process \citep{kelly2009variations} is defined as
\begin{equation}
d X(t) = -\frac{1}{\tau}({X}(t) - {\mu})dt + {\sigma} d {B}(t),\nonumber
\end{equation}
where $X(t)$ denotes the  magnitude of an astronomical object at time $t\in\mathbb{R}$, $\tau$ is the timescale of the process in days, $\mu$ is the long-term average magnitude of the process, $\sigma$ is the short-term variability of the process on the magnitude scale, and $B(t)$ is the standard Brownian motion.

A multivariate version of the damped random walk process \citep{gardiner2009} is defined as
\begin{equation}
d \boldsymbol{X}(t) = -D_{\boldsymbol{\tau}}^{-1}(\boldsymbol{X}(t) - \boldsymbol{\mu})dt + D_{\boldsymbol{\sigma}} d \boldsymbol{B}(t),\label{mOU}
\end{equation}
where  $\boldsymbol{X}(t)=\{X_1(t), \ldots, X_k(t)\}$ is a vector of length $k$ that denotes magnitudes of the $k$ bands at time $t\in\mathbb{R}$, ${D}_{\boldsymbol{\tau}}=\textrm{diag}(\tau_1, \ldots, \tau_k)$ is a $k\times k$ diagonal matrix whose diagonal elements are $k$  timescales  with each $\tau_j$ representing the  timescale of the $j$-th band in days, $\boldsymbol{\mu}=\{\mu_1, \ldots, \mu_k\}$ is a vector for long-term average magnitudes of $k$ bands, and $D_{\boldsymbol{\sigma}}=\textrm{diag}(\sigma_1, \ldots, \sigma_k)$ is a  $k\times k$ diagonal matrix whose diagonal elements are short-term variabilities (in magnitudes) of $k$ bands.  Finally, $\boldsymbol{B}(t)=\{B_1(t), \ldots, B_k(t)\}$ is a vector for $k$ standard Brownian motions whose pairwise correlations are modeled by correlation  parameters $\rho_{jl}~(1\le j<l\le k)$ such that $dB_j(t)B_l(t) = \rho_{jl} dt$. These correlations are essentially cross-correlations because they govern the correlations among different continuous-time processes.  (The subscripts $j$ and $l$ will be numeric hereafter, i.e., the subscripts $j$ and $l$ denote the $j$-th band and $l$-th band, respectively, not $j$-band and $l$-band.) We use $\boldsymbol{\rho}$ to denote a vector of these $k(k-1)/2$ correlation  parameters.

\begin{figure}
\begin{center}
\includegraphics[scale = 0.29]{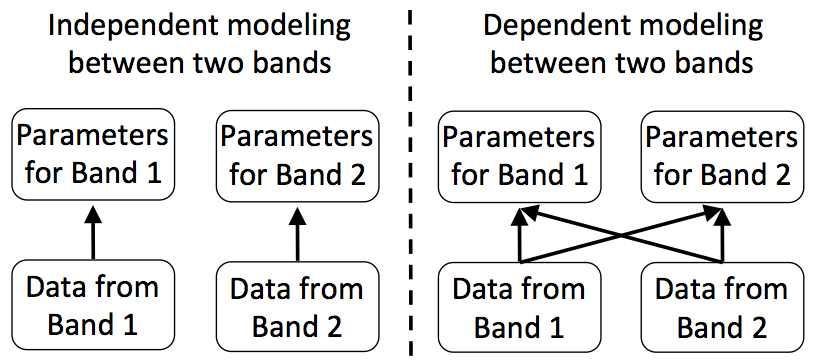}
\caption{A diagram illustrating the advantage of jointly modeling multi-band data. The independent assumption (zero correlation/covariance in a GP) between two bands makes it impossible for the data from one band to affect the parameter estimation of the other band. However, a dependent model between two bands with a correlation parameter enables the entire data from all bands to contribute to the parameter estimation of all bands. \label{fig1illu}}
\end{center}
\end{figure}

These correlation parameters are the key to the multi-band modeling. If these correlations are set to zeros, then the proposed multi-band model is essentially equivalent to a single-band model with an independent assumption across bands. In this case, data from one band do not contribute to the parameter estimation of other bands; see the left panel of Figure~\ref{fig1illu}.  This is because a zero correlation between two multivariate Gaussian random variables \textcolor{black}{implies} their independence. The proposed multi-band model accounts for the dependence between bands by introducing their correlation parameters. \textcolor{black}{This}  enables sharing information across multi-band data to infer parameters of all bands;  see the right panel of Figure~\ref{fig1illu}. 


\textcolor{black}{The matrix $D_{\tau}$ in \eqref{mOU} does not need to be diagonal as  long as it is positive definite \citep{vatiwutipong2019}. But here we limit it to be diagonal for better interpretability and computational efficiency.  If $D_{\tau}$ is a general positive definite matrix,  timescales become correlated via off-diagonal elements of $D_{\tau}$. One downside is that it is difficult to interpret  off-diagonal elements of $D_{\tau}$.  If $D_{\tau}$ is diagonal,  on the other hand, it is clear that $\tau_i$ is interpreted as the timescale of the $i$-th band. Such simple interpretability makes a simulation of the proposed more intuitive because it is enough to specify $k$ timescales instead of filling out off-diagonal elements of $D_{\tau}$. Second, if $D_{\tau}$ is diagonal, we can easily calculate a matrix exponent, e.g., $\exp(-(t-s)D_{\tau}^{-1})$ in~\eqref{dist:latent} below, significantly reducing computational burden with less parameters.}

The solution of the stochastic differential equation in~\eqref{mOU} is Gaussian, Markovian, and stationary \citep[105p,][]{gardiner2009}, i.e., given  $\boldsymbol{X}(s)$ and for $t\ge s$,
\begin{align}
\begin{aligned}\label{dist:latent}
&\boldsymbol{X}(t)\mid \boldsymbol{X}(s), \boldsymbol{\mu},  \boldsymbol{\sigma}, \boldsymbol{\tau}, \boldsymbol{\rho}\\
&\sim \mathrm{MVN}_k(\boldsymbol{\mu} + e^{-(t-s)D_{\boldsymbol{\tau}}^{-1}}(\boldsymbol{X}(s) - \boldsymbol{\mu}),~ Q(t-s)),
\end{aligned}
\end{align}
where MVN$_k(a, b$) represents a $k$-dimensional multivariate Gaussian distribution with mean vector $a$ and covariance matrix $b$. The $(j, l)$ entry of the covariance matrix $Q(t-s)$ is defined as
\begin{equation}
q_{jl} = \frac{\sigma_j\sigma_l\rho_{jl}\tau_j\tau_l}{\tau_j + \tau_l}\left(1 - e^{-(t-s)\frac{\tau_j+\tau_l}{\tau_j\tau_l}}\right).\label{qmatrix}
\end{equation}
We evaluate this continuous-time process at $n$ discrete observation times $\boldsymbol{t}=\{t_1, \ldots, t_n\}$. Then the joint probability density function of $\boldsymbol{X}(\boldsymbol{t})=\{\boldsymbol{X}(t_1), \ldots, \boldsymbol{X}(t_n)\}$ is
\begin{align}
\begin{aligned}\label{pdf:latent}
&f_1(\boldsymbol{X}(\boldsymbol{t}) \mid \boldsymbol{\mu},  \boldsymbol{\sigma}, \boldsymbol{\tau}, \boldsymbol{\rho})\\
& = \prod_{i=1}^n f_2(\boldsymbol{X}(t_i) \mid \boldsymbol{X}(t_{i-1}), \boldsymbol{\mu},  \boldsymbol{\sigma}, \boldsymbol{\tau}, \boldsymbol{\rho}),
\end{aligned}
\end{align}
where $f_2$ denotes the density function of the multivariate Gaussian distribution defined in \eqref{dist:latent},  $t_0 = -\infty$, and the subscript $i$ will be used to distinguish observation times ($i=1, 2, \ldots, n$). 



The observed data $\boldsymbol{x}=\{\boldsymbol{x}_1, \ldots, \boldsymbol{x}_n\}$ are multi-filter light curves measured at irregularly spaced observation times $\boldsymbol{t}$ with known  measurement error standard deviations, $\boldsymbol{\delta}=\{\boldsymbol{\delta}_1,  \ldots, \boldsymbol{\delta}_n\}$. Since one or more bands can be used at each observation time $t_i$, the length of a vector $\boldsymbol{x}_i$  can be different, depending on how many bands are used at the $i$-th observation time. For example, if  $g$- and $r$-bands are used at time $t_i$, then $\boldsymbol{x}_i$ is a vector containing  two magnitudes from the $g$- and $r$-bands, and $\boldsymbol{\delta}_i$ is a vector of  two corresponding measurement error standard deviations. We assume that these observed data are realizations of the latent  multi-filter light curves $\boldsymbol{X}(\boldsymbol{t})=\{\boldsymbol{X}(t_1), \ldots, \boldsymbol{X}(t_n)\}$  with known Gaussian measurement error variances $\boldsymbol{\delta}^2$. That is, for $i=1, \ldots, n$, 
\begin{equation}
\boldsymbol{x}_i\mid \boldsymbol{X}(t_i)\sim \textrm{MVN}_{k_i}(\boldsymbol{X}^\ast(t_i),~ D_{\boldsymbol{\delta}_i}^2),\label{dist:obs}
\end{equation}
where $k_i$ is the number of bands used at observation time $t_i$, and $\boldsymbol{X}^\ast(t_i)$ denotes a sub-vector of $\boldsymbol{X}(t_i)$ corresponding to the bands that are used to observe $\boldsymbol{x}_i$.  For example, if $g$- and $r$-bands are used for measuring $\boldsymbol{x}_i$ at $t_i$, then $\boldsymbol{X}^\ast(t_i)$ is a vector of length two ($k_i=2$) composed of the two elements of $\boldsymbol{X}(t_i)$ corresponding to the $g$- and $r$-bands. The notation $D_{\boldsymbol{\delta}_i}$ denotes a $k_i\times k_i$ diagonal matrix whose diagonal elements are $\boldsymbol{\delta}_i=\{\delta_{i1}, \ldots, \delta_{ik_i}\}$. These observed data $\boldsymbol{x}$ are assumed to be conditionally independent given the latent data $\boldsymbol{X}(\boldsymbol{t})$. \textcolor{black}{Thus} the resulting joint probability density function of the observed data given the latent data is expressed by
\begin{equation}
h_1(\boldsymbol{x}\mid \boldsymbol{X}(\boldsymbol{t})) = \prod_{i=1}^n h_2(\boldsymbol{x}_i\mid\boldsymbol{X}(t_i))\label{pdf:obs},
\end{equation}
where $h_2$ is the multivariate Gaussian density function defined in~\eqref{dist:obs}.

We summarize the proposed state-space representation in the following diagram.  

\begin{tikzpicture}
  \matrix (m) [matrix of math nodes,row sep=2em,column sep=1em]
  {
    \textrm{State}: &  \boldsymbol{X}(t_1) & \boldsymbol{X}(t_2) & \cdots &
      \boldsymbol{X}(t_n) \\
    \textrm{Space}: & \boldsymbol{x_1} & \boldsymbol{x_2} & ~ & \boldsymbol{x_n}\\};
  \path[-stealth]
    (m-1-2) edge  node [left] {} (m-1-3)
            edge  node [below] {} (m-2-2)
    (m-1-3) edge  node [left] {} (m-1-4)
            edge  node [below] {} (m-2-3)
    (m-1-4) edge  node [left] {} (m-1-5)
    (m-1-5) edge  node [below] {} (m-2-5);
\end{tikzpicture}

\noindent The arrows represent dependent and conditionally independent relationships. For example, both $\boldsymbol{X}(t_2)$ and $\boldsymbol{x_1}$ depend only on $\boldsymbol{X}(t_1)$, and they are conditionally independent given $\boldsymbol{X}(t_1)$ because there is no direct arrow between $\boldsymbol{X}(t_2)$ and $\boldsymbol{x_1}$. The conditional distributions of the latent magnitudes in the State level are defined in~\eqref{dist:latent}, and those of the observed data given the latent magnitudes in the space level are given in~\eqref{dist:obs}. The advantage of this state-space approach is that we can model the noisy observations $\boldsymbol{x}$ with known measurement error variances $\boldsymbol{\delta}^2$, as is done in~\eqref{dist:obs}. See also \cite{kelly2009variations} and \cite{kelly2014flexible} for the state-space representations of univariate CARMA(1, 0) and CARMA($p, q$), respectively, and Sec.~3 of \cite{durbin2012time} for details of state-space representation of GPs. 

Consequently, the  likelihood function of the model parameters with the latent process integrated out is
\begin{align}
\begin{aligned}\label{lik}
&L(\boldsymbol{\mu},  \boldsymbol{\sigma}, \boldsymbol{\tau}, \boldsymbol{\rho})\\
&=\int h_1(\boldsymbol{x}\mid \boldsymbol{X}(\boldsymbol{t}))f_1(\boldsymbol{X}(\boldsymbol{t}) \mid \boldsymbol{\mu},  \boldsymbol{\sigma}, \boldsymbol{\tau}, \boldsymbol{\rho}) ~d\boldsymbol{X}(\boldsymbol{t}).
\end{aligned}
\end{align}
Here $f_1$ and $h_1$ are defined in~\eqref{pdf:latent} and \eqref{pdf:obs}, respectively.

\section{Computation of the likelihood  function via Kalman-filtering}\label{sec3}
Kalman-filtering \citep{kalman1960} is a well-known technique  to evaluate the likelihood function of a state-space model when both state and space models are Gaussian.  \textcolor{black}{The} proposed \textcolor{black}{process} has a Gaussian state-space representation as shown in~\eqref{dist:latent} and \eqref{dist:obs}. \textcolor{black}{Thus,} it is natural to adopt  Kalman-filtering  to compute the likelihood function in~\eqref{lik} via a product of $n$ $k_i$-dimensional multivariate Gaussian densities ($i=1, \ldots, n$). This leads to $O(\sum_{i=1}^nk_i^3)$ complexity. The minimum complexity is $O(n)$ when only one band is used at each observation time ($k_i=1$) and the maximum complexity is $O(nk^3)$ when all of the $k$ bands are used  at every observation time ($k_i=k$). 
Let $\mathcal{F}(t_i)$ denote the natural filtration at time $t_i$, i.e., all of the information about the observed data available until time $t_i$. Using this notation,  we define the following predictive mean vector and covariance matrix at $t_{i-1}$: with $\Delta t_i=t_i-t_{i-1}$,
\begin{align}
\begin{aligned}\label{cond1}
\boldsymbol{\mu}_{i\vert i-1} &= E(\boldsymbol{X}(t_i)\mid \mathcal{F}(t_{i-1}), \boldsymbol{\mu},  \boldsymbol{\sigma}, \boldsymbol{\tau}, \boldsymbol{\rho})\\
&=\boldsymbol{\mu} + e^{-\Delta t_i D_{\boldsymbol{\tau}}^{-1}}(\boldsymbol{\mu}_{i-1\vert i-1} - \boldsymbol{\mu}),\\
\Sigma_{i\vert i-1} &= \textrm{Cov}(\boldsymbol{X}(t_i)\mid \mathcal{F}(t_{i-1}), \boldsymbol{\mu},  \boldsymbol{\sigma}, \boldsymbol{\tau}, \boldsymbol{\rho})\\
&=e^{-\Delta t_i D_{\boldsymbol{\tau}}^{-1}} (\Sigma_{i-1\vert i-1})e^{-\Delta t_i D_{\boldsymbol{\tau}}^{-1}} + Q(\Delta t_i).
\end{aligned}
\end{align}
We assume that
\begin{equation}\nonumber
\boldsymbol{\mu}_{1\vert0} = \boldsymbol{\mu}~~\textrm{and}~~\Sigma_{1|0} = \{q_{jl}\}= \left\{\frac{\sigma_j\sigma_l\rho_{jl}\tau_j\tau_l}{\tau_j + \tau_l}\right\}.
\end{equation}
Here, each element of the covariance matrix $Q(\Delta t_i)$ in~\eqref{cond1} is defined in~\eqref{qmatrix}, and the updated mean vector and covariance matrix (after observing data at $t_i$) are
\begin{align}
\begin{aligned}\nonumber
\boldsymbol{\mu}_{i\vert i} &= E(\boldsymbol{X}(t_i)\mid \mathcal{F}(t_{i}), \boldsymbol{\mu},  \boldsymbol{\sigma}, \boldsymbol{\tau}, \boldsymbol{\rho})\\
&=\boldsymbol{\mu}_{i\vert i-1} + \Sigma_{i|i-1}^{., \ast}(\Sigma^{\ast, \ast}_{i|i-1} + D^2_{\boldsymbol{\delta}_i})^{-1}(\boldsymbol{x}_i - \boldsymbol{\mu}_{i|i-1}), \\
\Sigma_{i|i} &= \textrm{Cov}(\boldsymbol{X}(t_i)\mid \mathcal{F}(t_{i}), \boldsymbol{\mu},  \boldsymbol{\sigma}, \boldsymbol{\tau}, \boldsymbol{\rho})\\
&=\Sigma_{i|i-1} - \Sigma^{., \ast}_{i|i-1}(\Sigma^{\ast, \ast}_{i|i-1} + D^2_{\boldsymbol{\delta}_i})^{-1}\Sigma^{\ast, .}_{i|i-1}.
\end{aligned}
\end{align}
The notation $\Sigma^{\ast, \ast}_{i\vert i-1}$ denotes a sub-matrix of $\Sigma_{i\vert i-1}$ restricted to the  bands used for observing $\boldsymbol{x}_i$,  $\Sigma^{\ast,.}_{i|i-1}$ is a sub-matrix of $\Sigma_{i|i-1}$ whose rows correspond to  the  bands used and columns correspond to  the entire bands, and  $\Sigma^{.,\ast}_{i|i-1}$ is a sub-matrix of $\Sigma_{i|i-1}$ whose rows correspond to  the  entire bands and columns correspond to the  bands used. For example, suppose there are five bands, $u, g, r, i, z$, and we use $u$- and $r$-bands to observe $\boldsymbol{x}_i$. Then, $\Sigma^{\ast, \ast}_{i\vert i-1}$ is a 2$\times$2 covariance matrix constructed by selecting rows and columns corresponding to $u$- and $r$-bands from $\Sigma_{i\vert i-1}$,  $\Sigma^{\ast, .}_{i\vert i-1}$ is a  2$\times$5 matrix made by choosing two rows corresponding to $u$- and $r$-bands and all columns from $\Sigma_{i\vert i-1}$, and  $\Sigma^{., \ast}_{i\vert i-1}$ is a  5$\times$2 matrix built by choosing all rows and two columns corresponding to $u$- and $r$-bands from $\Sigma_{i\vert i-1}$.

Consequently, the  likelihood function in~\eqref{lik} can be computed as follows:
\begin{equation}
L(\boldsymbol{\mu},  \boldsymbol{\sigma}, \boldsymbol{\tau}, \boldsymbol{\rho}) = \prod_{i=1}^n p(\boldsymbol{x}_i\mid \mathcal{F}(t_{i-1}), \boldsymbol{\mu},  \boldsymbol{\sigma}, \boldsymbol{\tau}, \boldsymbol{\rho}),\label{lik.kalman}
\end{equation}
where $p$ is another multivariate Gaussian  density of
\begin{align}
\begin{aligned}\label{lik.kalman.detail}
&\boldsymbol{x}_i\mid \mathcal{F}(t_{i-1}), \boldsymbol{\mu},  \boldsymbol{\sigma}, \boldsymbol{\tau}, \boldsymbol{\rho}\\
&\sim \textrm{MVN}_{k_i}(\boldsymbol{\mu}^\ast_{i\vert i-1},~ \Sigma^{\ast, \ast}_{i|i-1} + D^2_{\boldsymbol{\delta}_i}). 
\end{aligned}
\end{align}
The notation $\boldsymbol{\mu}^\ast_{i\vert i-1}$  denotes  a sub-vector of $\boldsymbol{\mu}_{i\vert i-1}$ restricted only to the bands used to observe $\boldsymbol{x}_i$.


By definition, the maximum likelihood estimates of the model parameters are the values that jointly maximize $L(\boldsymbol{\mu},  \boldsymbol{\sigma}, \boldsymbol{\tau}, \boldsymbol{\rho})$. We use a gradient-free optimization algorithm \citep{nelder1965} to obtain the maximum likelihood estimates.



\section{Bayesian inference}\label{sec4}

For  Bayesian hierarchical modeling, we adopt scientifically motivated, weakly informative, and independent prior distributions on the model parameters \citep{tak2017bayesian, tak2018how}: for $j=1, 2, \ldots, k$ and $j<l\le k$,
\begin{align}
\begin{aligned}\label{prior}
\mu_j&\sim \textrm{Unif}(-30,~ 30), &\tau_j\sim \textrm{inv-Gamma}(1, 1),\\
\rho_{jl}&\sim \textrm{Unif}(-1,~ 1),&\sigma_j^2\sim \textrm{inv-Gamma}(1, c),
\end{aligned}
\end{align}
where inv-Gamma$(a, b)$ denotes the inverse-Gamma distribution with shape parameter $a$ and scale parameter~$b$. We assume that each long-term average magnitude $\mu_j$ is in a reasonably wide range between $-30$ and 30. The correlation parameters are between $-1$ and 1 by definition. Setting up an  inverse-Gamma$(a, b)$ prior on an unknown parameter is considered as setting up a soft lower bound, $a/(b+1)$, of the unknown parameter \citep[Sec.~4.2, ][]{tak2018how}. We set up soft lower bounds of $\tau_j$'s and $\sigma^2_j$'s to prevent undesirable limiting behaviors of a damped random walk process \citep[Sec.~2.5, ][]{tak2017bayesian}. Specifically, the observed time series will look like a white-noise process if the timescale of the process is much smaller than the typical observation cadence (i.e., in the limit of timescale going to zero). Thus, we set up a half-day soft lower bound for each $\tau_j$. This undesirable behavior is also expected if the short-term variability (variance) of the process is much smaller than the typical measurement error variance. To reflect this, the constant $c$ in the inverse-Gamma prior of $\sigma^2_j$ in~\eqref{prior} is set to an arbitrarily small constant,  $10^{-7}$, so that the soft lower bound of each $\sigma_j$ is 0.00022.

Let $q$ be a joint prior density function of $\boldsymbol{\mu}$, $\boldsymbol{\sigma}$, $\boldsymbol{\tau}$, and $\boldsymbol{\rho}$ whose distributions are specified in~\eqref{prior}. Then, the resulting  full posterior density function $\pi$ is 
\begin{align}\label{full.post}
\pi(\boldsymbol{\mu},  \boldsymbol{\sigma}, \boldsymbol{\tau}, \boldsymbol{\rho} \mid \boldsymbol{x})\propto L(\boldsymbol{\mu},  \boldsymbol{\sigma}, \boldsymbol{\tau}, \boldsymbol{\rho})\times q(\boldsymbol{\mu},  \boldsymbol{\sigma}, \boldsymbol{\tau}, \boldsymbol{\rho}).
\end{align}

We adopt a Metropolis-Hastings within Gibbs sampler  \citep{tierney1994markov} to draw (dependent) posterior samples from the full posterior distribution $\pi(\boldsymbol{\mu},  \boldsymbol{\sigma}, \boldsymbol{\tau}, \boldsymbol{\rho} \mid \boldsymbol{x})$. Initial values of the model parameters are set to their maximum likelihood estimates. Then, it sequentially updates each parameter given the observed data and all the other parameters at each iteration. For example, suppose we have three parameters $\theta_1$, $\theta_2$, and $\theta_3$ to be updated at iteration~$s$ given previously updated values. We sequentially update each parameter as follows:
\begin{align}
&\textrm{Given}~~\left(\theta_1^{(s-1)}, \theta_2^{(s-1)}, \theta_3^{(s-1)}\right),\nonumber\\
&\textrm{sample}~\pi_1\!\left(\theta_1\mid \theta_2^{(s-1)}, \theta_3^{(s-1)}, \boldsymbol{x}\right),~\textrm{setting it to}~\theta_1^{(s)},\nonumber\\
&\textrm{sample}~\pi_2\!\left(\theta_2\mid \theta_1^{(s)}, \theta_3^{(s-1)}, \boldsymbol{x}\right),~\textrm{setting it to}~\theta_2^{(s)},\nonumber\\
&\textrm{sample}~\pi_2\!\left(\theta_3\mid \theta_1^{(s)}, \theta_2^{(s)}, \boldsymbol{x}\right),~\textrm{setting it to}~\theta_3^{(s)}.\nonumber
\end{align}
The parenthesized superscript indicates at which iteration the value of the parameter is updated. We use a truncated Gaussian  distribution between $-30$ and $30$ as a proposal distribution for each $\mu_j$, a truncated Gaussian proposal distribution between $-1$ and $1$ for each  $\rho_{jl}$, and a log-Normal proposal distribution for each of $\sigma_j^2$ and  $\tau_j$. We also use an adaptive Markov chain Monte Carlo  \citep[Sec.~3.2, ][]{tak2017bayesian} so that proposal scales  (or so-called jumping scales) are automatically tuned. 

 A tuning-free R package, \texttt{Rdrw}, that can analyze and simulate  both single- and multi-band light curves according to the proposed model is publicly available at CRAN\footnote{https://cran.r-project.org/package=Rdrw}.

\section{Numerical illustrations}\label{sec5}

\subsection{A simulation study on five-band light curves}\label{sec5.1}



We simulate a set of five-band light curves from the proposed multi-band model via a two-step procedure. The first step simulates \textcolor{black}{complete} data and the second step deletes some of them to be more realistic, e.g., making seasonal gaps. The simulation setting of the first step is as follows: (i) the total number of unique observation times across five bands is  set to 300; (ii) the observation cadences are randomly drawn from the Gamma($\alpha=3,~\beta=1$) distribution whose mean \textcolor{black}{is} 3 days; (iii) the measurement error standard deviations of the $j$-th band are randomly drawn from the $N(0.01 +0.004(j-1),~0.002^2)$ distribution ($j=1, \ldots, 5$) so that the first band accompanies the smallest measurement error standard deviations and the last band involves the largest;   (iv)  the generative parameter values are $\mu_j=17+0.5(j-1)$, $\sigma_j=0.01j$, $\tau_j=100+20  j$ for $j=1, \ldots, 5$. Also, $\rho_{jl}=1.1^{-\vert j-l\vert}$  for $1\le j< l\le 5$ so that the first and last bands are least correlated; (v) given these parameter values, the latent magnitudes $\boldsymbol{X}(\boldsymbol{{t}})$ are generated via~\eqref{dist:latent}; (vi) finally, conditioning on these latent magnitudes, the observed magnitudes $\boldsymbol{x}$ are generated via~\eqref{dist:obs}. The R package \texttt{Rdrw} has a functionality to return $\boldsymbol{x}$ as a $300\times5$ matrix.

\begin{figure}
\begin{center}
\includegraphics[scale = 0.29]{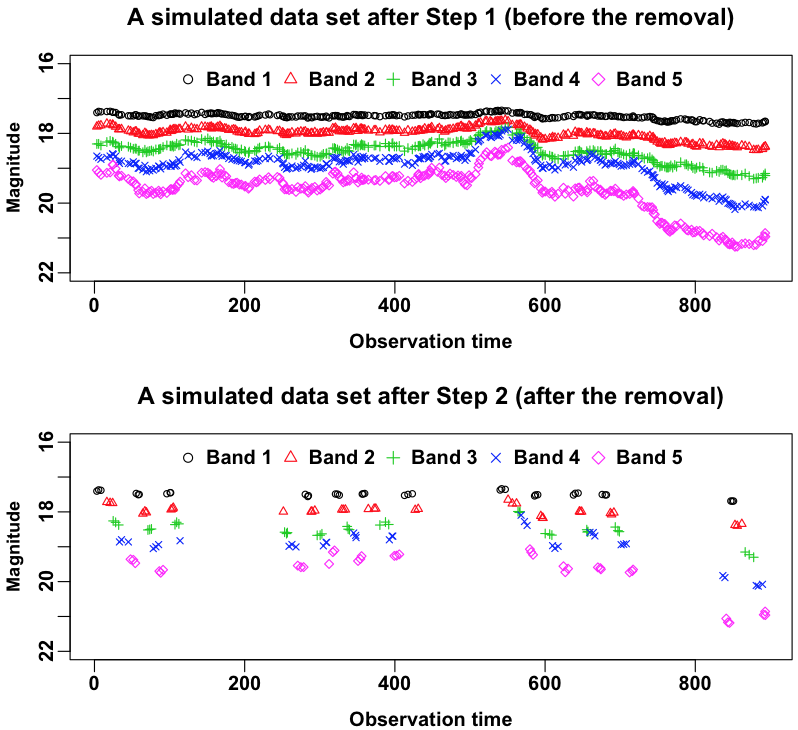}
\caption{A simulated data set of five-band light curves before (top) and after (bottom) removing some observations. The deleted data are based on three assumptions: (i) there is only a single-band measurement at each observation time; (ii) three consecutive observations are made for each band; (iii)  three seasonal gaps exist. After the deletion, the total number observation times is 180, and each band has 36 observations  with heteroscedastic measurement errors (whose standard deviations are too small to be displayed).\label{fig1sim}}
\end{center}
\end{figure}

Given these simulated data  in the first step, we remove some observations. \textcolor{black}{We} assume that only three consecutive observations occur for each band, and there are three seasonal gaps: (i) for the $j$-th band, we   keep only the following observation  numbers: $\{(1, 2, 3) + 3(j-1) + 15(b-1): b=1, 2, \ldots, 20\}$. For example, the observation numbers to be kept for the second band $(j=2)$  are 4, 5, 6, 19, 20, 21, $\ldots$, 289, 290, 291, and the other observations in the second band are removed. Similarly, the observation numbers for the third band ($j=3$) are 7, 8, 9, 22, 23, 24, $\ldots$, 292, 293, 294. By this rule, we can keep the 300 observation times in total and each observation time accompanies a  measurement from one band ($k_i=1$ for all $i$ in \eqref{lik.kalman.detail}, leading to $O(n)$ complexity). (ii) Next we create three seasonal gaps by removing 120 observation times out of 300. The observation numbers falling into the three seasonal gaps range from 41 to 80, from 141 to 180, and from 241 to 280. The total number of observations in the final data set is 180. The simulated five-band light curves before and after removing some of the data are displayed on the top and bottom panels of Figure~\ref{fig1sim}, respectively. The heteroscedastic measurement error standard deviations are too small to be displayed. Also, the details of the  simulated data are summarized in Table~\ref{table1sim}. 


Using this simulated data set (i.e., after the removal), we fit a univariate damped random walk model \citep{kelly2009variations} on each single-band light curve. And we  fit the proposed multivariate damped random walk model on the entire five-band data set. For each fit, we run a  Markov chain for 60,000 iterations, which is initiated at the maximum likelihood estimates of the model parameters. We discard the first 10,000 iterations as burn-in, and then we thin the remaining chain by a factor of five, i.e., from length 50,000 to 10,000. We use this thinned Markov chain Monte Carlo sample of size 10,000 to summarize the results. We check the convergence of the Markov chain by computing effective sample sizes  as a numerical indication of auto-correlation function; the higher the effective sample size is, the more quickly the auto-correlation function decreases.  The average effective sample size for the five short-term variabilities  is 2,896 and that for the five timescales is 1,866.

\begin{table}
	\centering
	\caption{The details of the simulated five-band time series data. We use `Med.'~to denote the median value, `mag.'~to indicate the magnitude value, and `SD' to represent the measurement error standard deviation. \textcolor{black}{Median cadence below is in days.} The first band data are the brightest with the smallest measurement uncertainty, and the last band data are the faintest with the largest measurement uncertainty. }
        \label{table1sim}
	\begin{tabular}{ccccc} 
		\hline
	Band	 & Length & Med.~cadence & Med.~mag. & Med.~SD \\
		\hline
$1$ & 36 & 3.368 & 17.492 &0.010\\
$2$ & 36 & 4.044 & 17.979 & 0.014\\
$3$  & 36 & 3.950 &18.513  & 0.018\\		
$4$ & 36 & 4.264 & 18.963 & 0.022\\
$5$  & 36 & 3.547 & 19.723  & 0.026\\
                 \hline
	\end{tabular}
\end{table}

\begin{figure}
\begin{center}
\includegraphics[scale = 0.28]{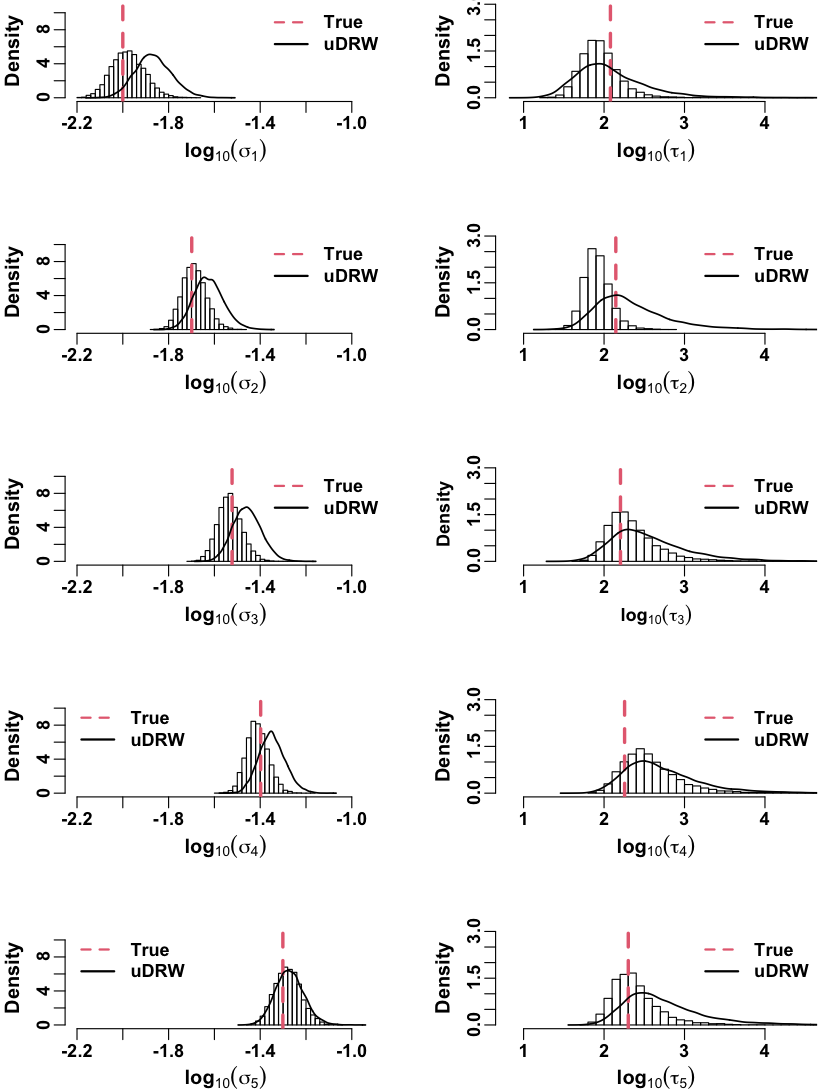}
\caption{The outcomes of fitting multi-band and single-band models on the simulated data. The five marginal posterior distributions (histograms) of the short-term variabilities obtained by the proposed multi-band model are displayed in the first column and those of the timescales are shown in the second column. The solid black curves represent the marginal posterior distributions of corresponding parameters obtained by the univariate damped random walk (uDRW) model. The vertical  dashed lines indicate the generative true parameter values. Overall, the marginal posterior distributions obtained by the multi-band model tend to have higher and narrower peaks near the generative true values.\label{fig2sim}}
\end{center}
\end{figure}


The marginal posterior distributions of the short-term variabilities obtained by the proposed multi-band model are exhibited in the first  column of Figure~\ref{fig2sim}.  \textcolor{black}{The second column of Figure~\ref{fig2sim} displays those of timescales.}  The black solid curves superimposed on the histograms denote the posterior distributions obtained by  the single-band model. The  vertical dashed lines represent the generative true values.  It turns out that the posterior distributions obtained by the proposed multi-band model (histograms) tend to have higher peaks and narrower spread around the generative true values than those obtained by the single-band models (solid curves). These results show that the proposed model results in more accurate inferences on the model parameters. This makes intuitive sense because the single-band model accesses only 36 observations in each band, while the multi-band model enables sharing information across bands via their correlations. That means,  the multi-band model allows all of the 180 observations to contribute to the inference on every model parameter \textcolor{black}{through their dependence}.

\begin{figure}
\begin{center}
\includegraphics[scale = 0.28]{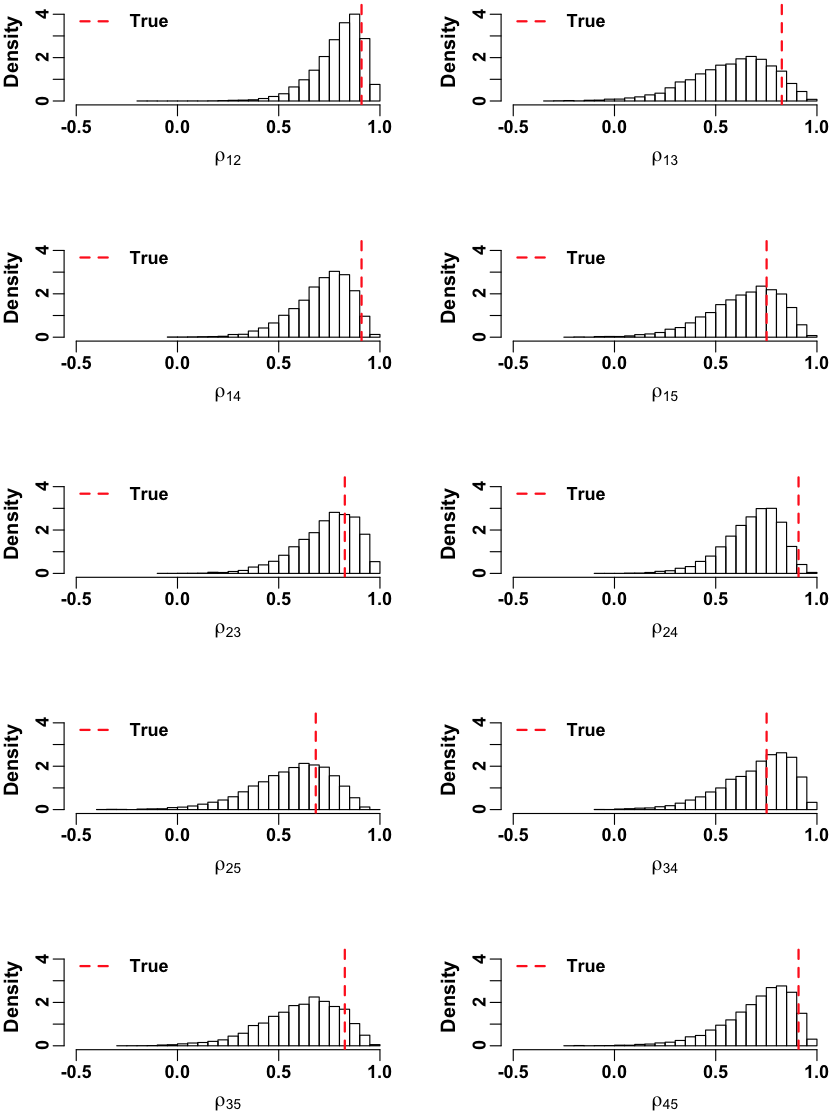}
\caption{The marginal posterior distributions of the ten cross-correlation parameters obtained by fitting the multi-band model on the simulated data set. The   vertical dashed lines indicate the generative true parameter values. It turns out that the marginal posterior distributions tend to have modes near the generative true values.\label{fig3sim}}
\end{center}
\end{figure}

When it comes to cross-correlation parameters, a comparison between single-band and multi-band models is not possible  because correlations are available only in the multi-band model. We display the marginal posterior distributions of the ten cross-correlation parameters in Figure~\ref{fig3sim}. It turns out that the true parameter values are closely located near the modes of these posterior distributions.

\subsection{Five-band light curves of an SDSS S82 quasar  }\label{sec5.2}
\begin{figure}
\begin{center}
\includegraphics[scale = 0.29]{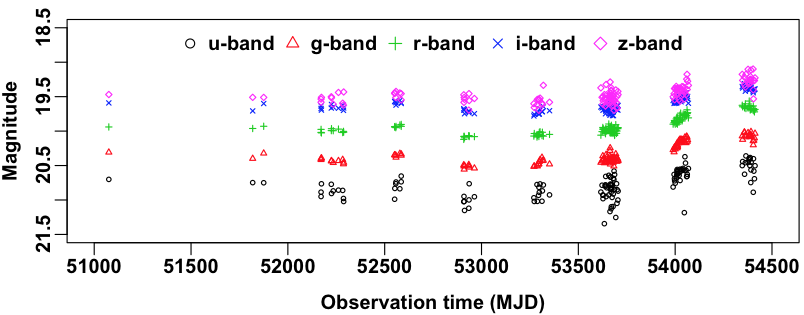}
\caption{The five-band light curves of a quasar 3078106 obtained from a catalog of 9,258 SDSS S82 quasars that are spectroscopically confirmed. Standard deviations  of heteroscedastic measurement errors are too small to be displayed. There is no tie in observation time, meaning that a single-band magnitude is measured at each observation time.\label{fig1}}
\end{center}
\end{figure}

The five-band ($u, g, r, i, z$) light curves of a quasar used in this illustration are obtained from a catalog of 9,258 SDSS Stripe 82 quasars that are spectroscopically confirmed \citep{macleod2012}\footnote{\url{http://faculty.washington.edu/ivezic/cmacleod/qso_dr7}}. The name of the quasar (dbID) is 3078106. We display these five-band light curves in Figure~\ref{fig1}. Most of the measurement error standard deviations are too small to be displayed in this figure.  There is no tie in observation times, i.e., a single-band magnitude is measured at each observation time ($k_i=1$ for all $i$ in \eqref{lik.kalman.detail}), leading to $O(n)$ complexity for the likelihood computation. The length of each single-band light curve is  different: 132 observations in $u$-band, 137 in $g$-band, 141 in $r$-band, 138 in $i$-band, and 139 in $z$-band. In total, there are 687  observation times across the bands ($n=687$); see  Table~\ref{table1} for more details of the data.

\begin{table}
	\centering
	\caption{Details of the multi-band time series data of a quasar 3078106. We use `Med.' to denote the median value, `mag.' to indicate the magnitude value, and `SD' to represent the measurement error standard deviation. \textcolor{black}{Median cadence below is in days.}}
        \label{table1}
	\begin{tabular}{ccccc} 
		\hline
	Band	 & Length & Med.~cadence & Med.~mag. & Med.~SD \\
		\hline
$u$ & 132 & 2.004& 20.768 &0.102\\
$g$ & 137 & 1.996 & 20.383 & 0.027\\
$r$  & 141 & 1.995 &19.995  & 0.024\\		
$i$  & 138 & 1.994 & 19.627 & 0.025\\
$z$  & 139 & 1.994 & 19.473  & 0.071\\
                 \hline
	\end{tabular}
\end{table}


We run a Markov chain of length 60,000, discarding the first 10,000 iterations as burn-in.  We thin the remaining chain by a factor of five from length 50,000 into 10,000.  We use this thinned Markov chain to make an inference on each parameter because the effective sample sizes are satisfactory across all model parameters. The sampling result is summarized in Table~\ref{table2}.

\begin{table}
	\centering
	\caption{Details of  the posterior samples of the model parameters. We use `Mean' to indicate the posterior mean, `SD' to represent the posterior standard deviation, and the 68.3\% credible interval is based on 15.85\% and 84.15\% quantiles of the corresponding posterior sample, and `ESS' to denote the effective sample size out of 10,000.  Numerical subscripts are indicators of the bands; 1, 2, 3, 4, and 5 correspond to $u, g, r, i$, and $z$, respectively.}
        \label{table2}
	\begin{tabular}{crrcr} 
		\hline
	Param.	 & Mean & SD &   68.3\% credible interval & ESS \\
		\hline
$\mu_1$  & 20.780 & 0.043 & (20.739, 20.820)& 379\\
$\mu_2$  & 20.347 & 0.036 & (20.312, 20.382)& 290\\
$\mu_3$   & 19.934 &0.034  & (19.902, 19.965)& 266\\		
$\mu_4$  & 19.616 & 0.029 &  (19.588, 19.644)& 380 \\
$\mu_5$  & 19.471 & 0.032  &  (19.442, 19.499)& 397\\
$\log_{10}(\sigma_1)$  & \textcolor{black}{$-1.887$} & \textcolor{black}{0.094} & \textcolor{black}{$(-1.983, -1.792)$} & 704 \\
$\log_{10}(\sigma_2)$  & \textcolor{black}{$-1.984$} & \textcolor{black}{0.064} & \textcolor{black}{$(-2.048, -1.919)$} & 697 \\
$\log_{10}(\sigma_3)$  & \textcolor{black}{$-2.060$} & \textcolor{black}{0.068} & \textcolor{black}{$(-2.129, -1.992)$} & 475 \\
$\log_{10}(\sigma_4)$  & \textcolor{black}{$-1.985$} & \textcolor{black}{0.071} & \textcolor{black}{$(-2.058, -1.913)$} & 804 \\
$\log_{10}(\sigma_5)$  & \textcolor{black}{$-1.992$} & \textcolor{black}{0.096} & \textcolor{black}{$(-2.088, -1.896)$} & 800 \\
$\log_{10}(\tau_1)$   & 2.147 & 0.212  & (1.937, 2.344)& 362\\
$\log_{10}(\tau_2)$   & 2.206 & 0.153  & (2.061, 2.350)& 326\\
$\log_{10}(\tau_3)$   & 2.253 & 0.175  & (2.087, 2.415)& 253\\
$\log_{10}(\tau_4)$   & 2.093 & 0.175  & (1.924, 2.260)& 337\\
$\log_{10}(\tau_5)$   & 2.097 & 0.212  & (1.898 2.296)& 430\\
$\rho_{12}$   & 0.811 & 0.135  & (0.685, 0.933)& 301\\
$\rho_{13}$  & 0.809 & 0.125 & (0.683, 0.929)& 476 \\
$\rho_{14}$  & 0.871 & 0.086  & (0.799, 0.944)& 493 \\
$\rho_{15}$  & 0.543 & 0.212  & (0.322, 0.759) & 199\\
$\rho_{23}$   & 0.740 & 0.123  & (0.632, 0.853)& 288\\
$\rho_{24}$  & 0.742 & 0.155  & (0.601, 0.882)& 220\\
$\rho_{25}$   & 0.490 & 0.238  & (0.239, 0.734)& 194\\
$\rho_{34}$   & 0.634 & 0.165  & (0.466, 0.795)& 287\\
$\rho_{35}$   & 0.686 & 0.182  & (0.503, 0.859)& 224\\
$\rho_{45}$  & 0.862 & 0.123  & (0.769, 0.959)& 493 \\
                 \hline
	\end{tabular}
\end{table}

\begin{figure}
\begin{center}
\includegraphics[scale = 0.28]{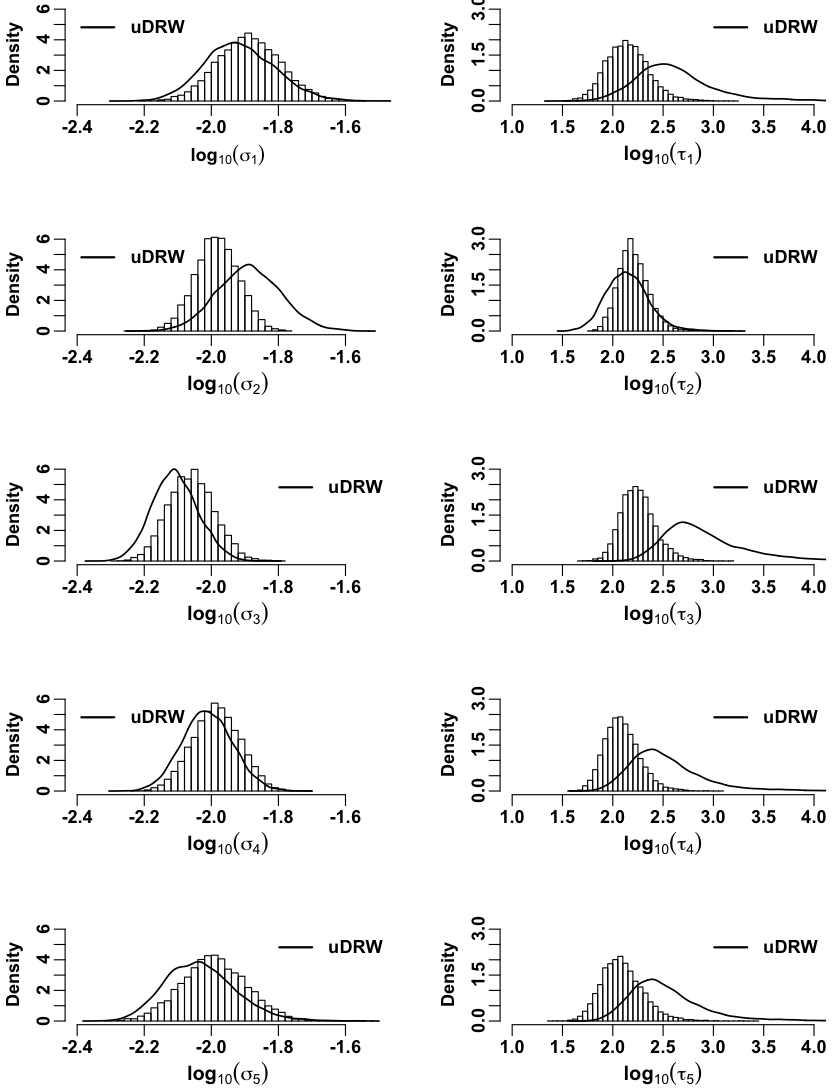}
\caption{The results of fitting multi-band and single-band models on the realistic data of SDSS S82 quasar 3078106. The marginal posterior distributions of the short-term variabilities obtained by the proposed multi-band model are displayed in the first column and those of the timescales are shown in the second column. The superimposed solid black curves  represent the marginal posterior distributions of corresponding parameters obtained by the univariate damped random walk (uDRW)  model. \textcolor{black}{Overall, the posterior distributions obtained by the multi-band model tend to have higher peak and narrower spread. Also, the modal locations of timescales in the second column are more consistent across bands under the multi-band model. This indicates a possibility that the timescales of this quasar are similar in all bands, which single-band models do not reveal.}  \label{fig4}}
\end{center}
\end{figure}

We compare the marginal posterior distributions of the short-term variabilities and timescales obtained by the proposed multi-band model with those obtained by the single-band models. Figure~\ref{fig4} displays these marginal posterior distributions of the short-term variabilities in the first column and those of the timescales in the second column. The histograms indicate the posterior distributions obtained by the multi-band model, while the superimposed solid black curves  represent those obtained by the single-band model. 

Overall, the marginal posterior distributions obtained by the multi-band model \textcolor{black}{(histograms)} tend to have higher peak and narrower spread than those obtained by the single-band models \textcolor{black}{(solid curves)}. In particular, the modes of all five marginal posterior distributions of timescales  in the second column are closely located near 2.3  \textcolor{black}{under the multi-band model}. \textcolor{black}{This can  be partially ascribed to a shrinkage effect of Bayesian hierarchical modeling \citep{efron1975data, morris1983parametric, tak2017rgbp}}. Single-band models do not exhibit such consistent modal locations. Thus, the result of multi-band modeling  indicates \textcolor{black}{a possibility} that the true timescales of quasar 3078106 may be \textcolor{black}{similar} across the five bands, \textcolor{black}{which} single-band analyses \textcolor{black}{do not reveal}.  

\subsection{Time delay estimation between doubly-lensed multi-band light curves of Q0957+561}\label{sec5.3}

Time delay estimation is one of the  key factors for the Hubble constant estimation via time delay cosmography \citep{kai2015, treu2016time, suyu2017holicow}. However, it has been the case that the time delays are estimated only from single-band light curves. The proposed  process  can be used to estimate time delays among gravitationally lensed multi-band light curves of an AGN, as is the case for a single-band model \citep{dobler2015, kai2015, tak2017bayesian}. For this purpose, we  introduce a few more parameters for the time delays and microlensing adjustment.

We use the proposed multi-filter process to model each of multiply lensed images observed in several bands, e.g., $\boldsymbol{X}_A(\boldsymbol{t})$,  $\boldsymbol{X}_B(\boldsymbol{t})$, $\boldsymbol{X}_C(\boldsymbol{t})$, and $\boldsymbol{X}_D(\boldsymbol{t})$ corresponding to quadruply lensed images, $A, B, C$, and $D$. (The notation here is consistent to the one in Section~\ref{modelspecification} except the subscripts that distinguish lensed images.) Taking lensed image $A$ as an example, the notation $\boldsymbol{X}_A(\boldsymbol{t})$ represents $\{\boldsymbol{X}_A(t_1), \ldots, \boldsymbol{X}_A(t_n)\}$ and each component $\boldsymbol{X}_A(t_i)$ is a vector of length $k$ (the number of available bands), i.e., $\boldsymbol{X}_A(t_i)=\{X_{A, 1}(t_i), \ldots, X_{A, k}(t_i)\}$.  The observed data of lensed image $A$ are $\boldsymbol{x}_A=\{\boldsymbol{x}_{A, 1}, \ldots, \boldsymbol{x}_{A, n}\}$ and corresponding measurement error standard deviations are $\boldsymbol{\delta}_{A}=\{\boldsymbol{\delta}_{A, 1}, \ldots, \boldsymbol{\delta}_{A, n}\}$.

We account for  microlensing by subtracting a polynomial long-term trend of order $m$ from each latent magnitude  \citep{hojjati2013robust, tewes2013a, tak2017bayesian}. That is,  for $j=1, \ldots, k$,
\begin{equation}
X'_{A, j}(t_i)=X_{A, j}(t_i)-p_{A, j}(t_i),
\end{equation}
where $p_{A, j}(t_i)$ is the polynomial long-term trend at $t_i$ defined as
\begin{equation}
p_{A, j}(t_i)=\beta_{A, j, 0}+\beta_{A, j, 1}t_i +\beta_{A, j, 2}t_i^2+\cdots+\beta_{A, j, m}t_i^m. 
\end{equation}
The notation $\boldsymbol{\beta}_{A, j}=\{\beta_{A, j, 0}, \beta_{A, j, 1}, \ldots, \beta_{A, j,  m}\}$ denotes the $m+1$ polynomial regression coefficients of band~$j$ for image~$A$. We use $\boldsymbol{\beta}_{A}=\{\boldsymbol{\beta}_{A, 1}, \ldots, \boldsymbol{\beta}_{A, k}\}$ to collectively denote the regression coefficients for image $A$. The polynomial order $m$ can be  set differently for each band \textcolor{black}{and for each lensed image} (e.g., $m_{A, 1}, \ldots, m_{A, k}$).  We note that the long-term average magnitude of band~$j$ for image $A$, i.e., $\mu_{A, j}$, is now absorbed into the intercept term of $\boldsymbol{\beta}_{A, j}$. 



We assume that each multi-band process, whose polynomial long-term trends are removed across $k$ bands, is a shifted version of the other in the horizontal axis by  the time delays. That means,
\begin{align}
\begin{aligned}
\boldsymbol{X}'_B(\boldsymbol{t})&=\boldsymbol{X}'_A(\boldsymbol{t}-\Delta_{AB}),\\\boldsymbol{X}'_C(\boldsymbol{t})&=\boldsymbol{X}'_A(\boldsymbol{t}-\Delta_{AC}),~ \textrm{and}\\\boldsymbol{X}'_D(\boldsymbol{t})&=\boldsymbol{X}'_A(\boldsymbol{t}-\Delta_{AD}).
\end{aligned}
\end{align}
Consequently,  given the time delays ($\Delta_{AB}$, $\Delta_{AC}$, $\Delta_{AD}$) and polynomial regression coefficients for microlensing ($\boldsymbol{\beta}_A$, $\boldsymbol{\beta}_B$, $\boldsymbol{\beta}_C$, $\boldsymbol{\beta}_D$), we can  use only one multi-band process $\boldsymbol{X}'_A(\cdot)$ to model all of the gravitationally lensed multi-filter light curves, i.e., $\boldsymbol{X}'_A(\boldsymbol{t})$, $\boldsymbol{X}'_A(\boldsymbol{t}-\Delta_{AB})$, $\boldsymbol{X}'_A(\boldsymbol{t}-\Delta_{AC})$, and  $\boldsymbol{X}'_A(\boldsymbol{t}-\Delta_{AD})$.

Moreover, given the time delays and polynomial regression coefficients for microlensing, we can unify the notation, combining all multi-band light curves of the lensed images. Let $\boldsymbol{\tilde{t}}=\{\tilde{t}_1, \ldots, \tilde{t}_{4n}\}$ be the sorted $4n$  observation times among $\boldsymbol{t}$, $\boldsymbol{t}-\Delta_{AB}$, $\boldsymbol{t}-\Delta_{AC}$, and $\boldsymbol{t}-\Delta_{AD}$. The unified notation for the observed data at $\tilde{t}_i$ is  $\boldsymbol{y}_i$  defined as follows: for $i=1, 2, \ldots, 4n$,
\begin{align}
\boldsymbol{y}_i = \left\{\begin{aligned}\nonumber
&\boldsymbol{x}_{A, i}~\textrm{}~\textrm{if}~~\tilde{t}_i\in \boldsymbol{t},\\
&\boldsymbol{x}_{B, i}~~\textrm{if}~~\tilde{t}_i\in \boldsymbol{t}-\Delta_{AB},\\
&\boldsymbol{x}_{C, i}~~\textrm{if}~~\tilde{t}_i\in \boldsymbol{t}-\Delta_{AC},\\
&\boldsymbol{x}_{D, i}~~\textrm{if}~~\tilde{t}_i\in \boldsymbol{t}-\Delta_{AD}\\
\end{aligned}\right.
\end{align}
with measurement error standard deviation
\begin{align}
\boldsymbol{\eta}_i = \left\{\begin{aligned}\nonumber
&\boldsymbol{\delta}_{A, i}~\textrm{}~\textrm{if}~~\tilde{t}_i\in \boldsymbol{t},\\
&\boldsymbol{\delta}_{B, i}~~\textrm{if}~~\tilde{t}_i\in \boldsymbol{t}-\Delta_{AB},\\
&\boldsymbol{\delta}_{C, i}~~\textrm{if}~~\tilde{t}_i\in \boldsymbol{t}-\Delta_{AC},\\
&\boldsymbol{\delta}_{D, i}~~\textrm{if}~~\tilde{t}_i\in \boldsymbol{t}-\Delta_{AD}.\\
\end{aligned}\right.
\end{align}
The unifying notation for the latent data $\boldsymbol{Y}(\tilde{t}_i)$ is
\begin{align}
\boldsymbol{Y}(\tilde{t}_i) = \left\{\begin{aligned}\nonumber
&\boldsymbol{X}'_A(\tilde{t}_i) + p_{A}(\tilde{t}_i)~~\textrm{if}~~\tilde{t}_i\in \boldsymbol{t},\\
&\boldsymbol{X}'_A(\tilde{t}_i) + p_{B}(\tilde{t}_i)~~\textrm{if}~~\tilde{t}_i\in \boldsymbol{t}-\Delta_{AB},\\
&\boldsymbol{X}'_A(\tilde{t}_i) + p_{C}(\tilde{t}_i)~~\textrm{if}~~\tilde{t}_i\in \boldsymbol{t}-\Delta_{AC},\\
&\boldsymbol{X}'_A(\tilde{t}_i) + p_{D}(\tilde{t}_i)~~\textrm{if}~~\tilde{t}_i\in \boldsymbol{t}-\Delta_{AD}\\
\end{aligned}\right.
\end{align}
where  $p_{A}(\tilde{t}_i)$ is a vector of length $k$ composed of $\{p_{A, 1 }(\tilde{t}_i), \ldots, p_{A, k}(\tilde{t}_i)\}$, and we similarly define $p_{B}(\tilde{t}_i)$, $p_{C}(\tilde{t}_i)$, and $p_{D}(\tilde{t}_i)$.

Using the unified notation, we  specify the  distributions for the proposed time delay model. Let us define $\boldsymbol{\Delta}=\{\Delta_{AB}, \Delta_{AC}, \Delta_{AD}\}$ and $\boldsymbol{\beta}=\{\boldsymbol{\beta}_A, \boldsymbol{\beta}_B, \boldsymbol{\beta}_C, \boldsymbol{\beta}_D\}$. Then the joint density function of the latent data is
\begin{align}
\begin{aligned}\label{td.unified1}
&f_1(\boldsymbol{Y}(\tilde{\boldsymbol{t}}) \mid  \boldsymbol{\sigma}, \boldsymbol{\tau}, \boldsymbol{\rho}, \boldsymbol{\Delta}, \boldsymbol{\beta})\\
& = \prod_{i=1}^{4n} f_2(\boldsymbol{Y}(\tilde{t}_i) \mid \boldsymbol{Y}(\tilde{t}_{i-1}), \boldsymbol{\sigma}, \boldsymbol{\tau}, \boldsymbol{\rho}, \boldsymbol{\Delta}, \boldsymbol{\beta}),
\end{aligned}
\end{align}
where the conditional distributions for $f_2$ are defined as
\begin{align}
\begin{aligned}\label{td.unified2}
&\boldsymbol{Y}(\tilde{t}_i)\mid \boldsymbol{Y}(\tilde{t}_{i-1}), \boldsymbol{\sigma}, \boldsymbol{\tau}, \boldsymbol{\rho}, \boldsymbol{\Delta}, \boldsymbol{\beta}\\
&\sim \mathrm{MVN}_k(e^{-(\tilde{t}_{i}-\tilde{t}_{i-1})D_{\boldsymbol{\tau}}^{-1}}\boldsymbol{Y}(\tilde{t}_{i-1}),~ Q(\tilde{t}_{i}-\tilde{t}_{i-1}))
\end{aligned}
\end{align}
for $i=1, 2, \ldots, 4n$. The $(j, l)$ entry of the covariance matrix $Q(\tilde{t}_{i}-\tilde{t}_{i-1})$ is defined in~\eqref{qmatrix}. The joint density function of the observed data is
\begin{equation}\label{td.unified3}
h_1(\boldsymbol{y}\mid \boldsymbol{Y}(\tilde{\boldsymbol{t}}), \boldsymbol{\Delta}, \boldsymbol{\beta}) = \prod_{i=1}^{4n} h_2(\boldsymbol{y}_i\mid\boldsymbol{Y}(\tilde{t}_i), \boldsymbol{\Delta}, \boldsymbol{\beta}),
\end{equation}
where the conditional distributions for $h_2$ are
\begin{equation}\label{td.unified4}
\boldsymbol{y}_{i}\mid \boldsymbol{Y}(\tilde{t}_i), \boldsymbol{\Delta}, \boldsymbol{\beta}\sim \textrm{MVN}_{k_i}(\boldsymbol{Y}^\ast(\tilde{t}_i),~ D_{\boldsymbol{\eta}_{i}}^2)
\end{equation}
for $i=1, 2, \ldots, 4n$.  The notation $\boldsymbol{Y}^\ast(\tilde{t}_i)$ denotes a sub-vector of $\boldsymbol{Y}(\tilde{t}_i)$ corresponding to the bands used for observing $\boldsymbol{y}_{i}$ at $\tilde{t}_i$. We also note that the conditions in~\eqref{td.unified1}--\eqref{td.unified4} include $\boldsymbol{\Delta}$ and $\boldsymbol{\beta}$  because without $\boldsymbol{\Delta}$ and $\boldsymbol{\beta}$ we cannot define $\boldsymbol{y}_{i}$'s and $\boldsymbol{Y}(\tilde{t}_i)$'s. The resulting likelihood function of the model parameters with the latent process marginalized out is
\begin{align}
\begin{aligned}\label{lik.td}
L(\boldsymbol{\sigma}, \boldsymbol{\tau}, \boldsymbol{\rho}, \boldsymbol{\Delta}, \boldsymbol{\beta})=&\int f_1(\boldsymbol{Y}(\tilde{\boldsymbol{t}}) \mid \boldsymbol{\sigma}, \boldsymbol{\tau}, \boldsymbol{\rho}, \boldsymbol{\Delta}, \boldsymbol{\beta})\\
& \times h_1(\boldsymbol{y}\mid \boldsymbol{Y}(\tilde{\boldsymbol{t}}), \boldsymbol{\Delta}, \boldsymbol{\beta})~d\boldsymbol{Y}(\tilde{\boldsymbol{t}}),
\end{aligned}
\end{align}
where $f_1$ and $h_1$ are defined in~\eqref{td.unified1} and \eqref{td.unified3}, respectively. The same Kalman-filtering procedure in Section~\ref{sec3} is used to calculate this likelihood function. The only difference is that there are two times more observations (from $n$ to $2n$)  for doubly-lensed system and four times more  (from $n$ to $4n$) for quadruply-lensed system. The number of unknown  parameters when $k$ bands are used for $a$ lensed images with the $m$-th order polynomial regression  is $2k+k(k-1)/2+(a-1)+ak(m+1)$. That is, $k$ $\sigma_j$'s, $k$ $\tau_j$'s, $k(k-1)/2$ $\rho_{jl}$'s, $a-1$ time delays, and $ak(m+1)$ polynomial regression coefficients. For example,  the number of unknown parameters is 103 when five bands are used ($k=5$) for a quadruply lensed quasar ($a=4$) with a cubic polynomial regression ($m=3$).

For a Bayesian inference, we adopt the same priors for $\boldsymbol{\sigma}$, $\boldsymbol{\tau}$, and $\boldsymbol{\rho}$ as specified in~\eqref{prior}, and weakly-informative independent prior distributions for the additional  parameters,  $\boldsymbol{\Delta}$ and $\boldsymbol{\beta}$ \citep[Sec.~2.4,][]{tak2017bayesian}. Specifically, we assume an independent multivariate Gaussian distribution, MVN$_{m+1}(\mathbf{0}_{m+1}, D)$, for each of $\boldsymbol{\beta}_{A, j}$'s, $\boldsymbol{\beta}_{B, j}$'s, $\boldsymbol{\beta}_{C, j}$'s, $\boldsymbol{\beta}_{D, j}$'s, where $\mathbf{0}_{m+1}$ denotes a vector of $m+1$ zeros and $D$ is a diagonal matrix whose diagonal elements are set to relatively large constants. We also adopt a Uniform distribution over the range between $\min(\tilde{\boldsymbol{t}})-\max(\tilde{\boldsymbol{t}})$ and $\max(\tilde{\boldsymbol{t}})-\min(\tilde{\boldsymbol{t}})$ for each of $\Delta_{BA}$, $\Delta_{CA}$, $\Delta_{DA}$.  We use $q(\boldsymbol{\sigma}, \boldsymbol{\tau}, \boldsymbol{\rho}, \boldsymbol{\Delta}, \boldsymbol{\beta})$ to denote the joint prior distribution of these unknown model parameters. The resulting full posterior distribution of the unknown model parameters is
\begin{equation}
\pi(\boldsymbol{\sigma}, \boldsymbol{\tau}, \boldsymbol{\rho}, \boldsymbol{\Delta}, \boldsymbol{\beta})\propto L(\boldsymbol{\sigma}, \boldsymbol{\tau}, \boldsymbol{\rho}, \boldsymbol{\Delta}, \boldsymbol{\beta})\times q(\boldsymbol{\sigma}, \boldsymbol{\tau}, \boldsymbol{\rho}, \boldsymbol{\Delta}, \boldsymbol{\beta}).
\end{equation}

We adopt the same adaptive Metropolis-Hastings within Gibbs sampler to draw posterior samples from this full posterior distribution. This time, it is natural to think of a two-step sampling scheme. First, given the parameters related to the time delay model, i.e., $\boldsymbol{\Delta}$ and $\boldsymbol{\beta}$, we can completely determine $\boldsymbol{y}$ and  $\boldsymbol{Y}(\tilde{\boldsymbol{t}})$. Thus, given $\boldsymbol{\Delta}$ and $\boldsymbol{\beta}$, we can update the parameters relevant to  the multivariate damped random walk process, i.e.,  $\boldsymbol{\sigma}$, $\boldsymbol{\tau}$, and $\boldsymbol{\rho}$, as done in Sections~\ref{sec5.1} and \ref{sec5.2}. Second, given the updated parameters, $\boldsymbol{\sigma}$, $\boldsymbol{\tau}$, and $\boldsymbol{\rho}$, we  update the time-delay-related parameters, $\boldsymbol{\Delta}$ and $\boldsymbol{\beta}$. Here we use a truncated Gaussian proposal distribution for each of $\boldsymbol{\Delta}$, and we do not need a proposal distribution for $\boldsymbol{\beta}$ because the conditional posterior distribution of $\boldsymbol{\beta}$ is a multivariate Gaussian.  The sampler iterates these two steps at each iteration. An  R package, \texttt{timedelay}, features this two-step update scheme to fit the proposed time delay model on a doubly lensed system observed in two bands\footnote{https://cran.r-project.org/package=timedelay}. (The package will be updated later for more general cases.)

\begin{figure}
\begin{center}
\includegraphics[scale = 0.387]{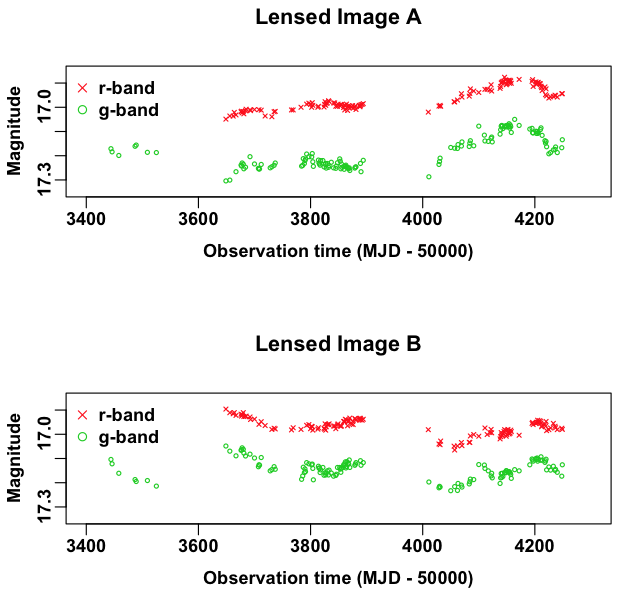}
\caption{Light curves of doubly-lensed quasar Q0957+561 \citep{refId0}. The $g$- and $r$-band light curves of lensed image $A$ appear in the top panel and those of lensed image $B$ appear in the bottom panel. Due to  strong gravitational lensing, multi-band light curves of one image lag behind by the time delay. The $g$-band light curve has more fluctuations, which is crucial in estimating the time delay.\label{fig1q0957}}
\end{center}
\end{figure}

Using the proposed time delay model and fitting procedure, we estimate the time delay between doubly lensed images of Q0957+561. The data are observed in $g$- and $r$-bands and are publicly available \citep{refId0}.   The details of the data are summarized in Table~\ref{q0957table}.  The $g$- and $r$-band light curves of lensed image $A$ are  displayed on the top panel of Figure~\ref{fig1q0957}, and those of lensed image $B$ are shown on the bottom panel of Figure~\ref{fig1q0957}. The resulting model involves 22 unknown model parameters.

\begin{table}
	\centering
	\caption{Details of the $r$- and $g$-band time series data of quasar Q0957+561 \citep{refId0}. We use `Med.' to denote the median value, `mag.' to indicate the magnitude value, and `SD' to represent the measurement error standard deviation.}
        \label{q0957table}
	\begin{tabular}{ccccc} 
		\hline
	Image	 & \multirow{2}{*}{Length} & \multirow{2}{*}{Med.~cadence} & \multirow{2}{*}{Med.~mag.} &  \multirow{2}{*}{Med.~SD} \\
		(Band)	 &  &  &  & \\
		\hline
$A~(r)$ & 132 & 1.978 & 16.983 &0.012\\
$A~(g)$ & 142 & 1.881 & 17.205 & 0.016\\
$B~(r)$  & 132 & 1.978 &16.966  & 0.012\\		
$B~(g)$  & 142 & 1.881 & 17.144 & 0.016\\
                 \hline
	\end{tabular}
\end{table}

To fit the proposed time delay model with a cubic polynomial regression for microlensing ($m=3$), we implement a Markov chain of length 60,000, discarding the first 10,000 iterations as burn-in. We thin the remaining chain by a factor of five (from 50,000 to 10,000). We summarize our inferential result using this thinned Markov chain. The effective sample size of the time delay $\Delta_{AB}$ is 675, leading to its auto-correlation quickly decreasing to zero. As for single-band fits, we adopt a quadratic polynomial regression ($m=2$) for the microlensing adjustment because a Markov chain with a cubic order ($m=3$) does not converge.

\begin{figure}
\begin{center}
\includegraphics[scale = 0.382]{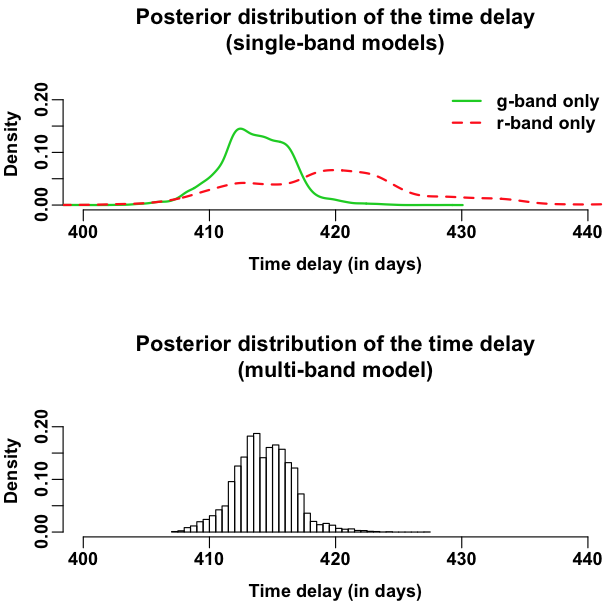}
\caption{The solid green  curve in the top panel represents the marginal posterior distribution of the time delay $\Delta_{AB}$ of Q0957+561 obtained by fitting the single-band model on the $g$-band data. Similarly the red dashed  curve in the top panel is the one from the $r$-band data. These single-band fits reveal a couple of possibilities for the time delay. In the bottom panel, however, the posterior distribution of $\Delta_{AB}$ obtained from the multi-band model  narrows down the possibility of the time delay by jointly modeling the two-band data.\label{fig1q0957.res}}
\end{center}
\end{figure}

The top panel of Figure~\ref{fig1q0957.res} shows the marginal posterior distributions of $\Delta_{AB}$  obtained by fitting two single-band models independently. The \textcolor{black}{solid} green curve indicates the marginal posterior distribution  of $\Delta_{AB}$ based only on the $g$-band data. \textcolor{black}{The resulting posterior mean of $\Delta_{AB}$  is 413.392 and its posterior standard deviation is 2.916}. The red dashed curve represents the distribution obtained  with only the $r$-band data \textcolor{black}{whose posterior mean is 420.597 and standard deviation is 8.165}. Clearly, the fitting results are not consistent. The $g$-band posterior distribution has a mode near 415 days, while the $r$-band posterior distribution shows two modes, one near 415 days and the other near 420 days. 

The bottom panel  of Figure~\ref{fig1q0957.res} exhibits the marginal posterior distribution of $\Delta_{AB}$  obtained by fitting the proposed multi-band model on the entire data. \textcolor{black}{The resulting posterior mean and standard deviation of $\Delta_{AB}$ are 414.324 and 2.307, respectively.}
It is now clear which mode the data support more. The relative height of the mode near 420 days is not even comparable to that of the mode near 415 days. \textcolor{black}{Also,} the peak near 415 days is higher \textcolor{black}{and narrower} than before\textcolor{black}{, which can be confirmed by the smaller posterior standard deviation}.  This makes intuitive sense because the $g$-band light curves have more fluctuations, as shown in Figure~\ref{fig1q0957}, and thus it is easier to find the time delay by matching those fluctuations. Consequently, the joint model  puts more weight on the information from   the $g$-band, enabling us to narrow the  two possibilities down to the highest mode near 415 days.


This feature of sharing information across different bands will be useful for finding time delays from the multi-band LSST light curves. This is because single-band LSST light curves are sparsely observed, and thus single-band observations may not exhibit enough fluctuation patterns \textcolor{black}{that are} needed to estimate time delays.


\section{Discussion}\label{sec6}

The proposed state-space representation of a multivariate damped random walk process can be applied to other sub-fields of astronomy as well. Photometric reverberation mapping is one  such \textcolor{black}{example}. \cite{zu2011an} model  continuum variability by a univariate GP with damped random walk covariance function to estimate AGN reverberation time lags. \cite{zu2016app} modify this model to infer a two-band reverberation time lag. They stack up two-band light curves in one vector and apply  the univariate GP framework with $O(n^3)$ complexity. We note that their two-band time-lag model is closely related to the single-band time delay model between doubly-lensed light curves. This is because the former reduces to the \textcolor{black}{latter} if an indicator function $I_{\Delta}(t)$, which is one when $t=\Delta$ and zero otherwise, is used as  a transfer function $\Psi(t)$. This implies that the proposed multivariate process  can be useful for improving their reverberating time-lag estimate, as is the case in multi-band time delay estimation. 
We also note that the proposed \textcolor{black}{model} can be a \textcolor{black}{parametric} alternative to a widely used non-parameteric \textcolor{black}{cross-correlation} method in multi-band data analyses, e.g., \cite{edelson2015space}. This is because the proposed multivariate process can be considered as a parametric cross-correlation method since it produces posterior distributions of all pair-wise cross-correlation parameters. Its applicability is not limited to correlations among  multi-band time series of one object. It can be used to find cross-correlations of multiple time series data of different sources if \textcolor{black}{their timescales are similar enough to make the cross-correlation  meaningful}.


\textcolor{black}{From a methodological point of view, the proposed model has several advantages over coregionalization \citep{journel1978, gelfand2004, alvarez2012kernels}, although the latter is more flexible. First, unlike the proposed damped random walk,  coregionalization  does not account for the \emph{damped} part. This is because it expresses each component of a vector-valued output as a linear function of independent GPs, which corresponds only to  the random walk part. Second, the proposed has better interpretability.  The dependence among multiple bands is clearly modeled by  well-known cross-correlation parameters $\rho_{ij}$'s. Also each light curve is modeled by a univariate damped random walk with only three parameters ($\mu_i, \sigma_i, \tau_i)$, i.e., long-term average, short-term variability, and timescale of a latent process.  On the other hand, the  coefficients of linear functions in coregionalization  do not have such intuitive interpretability. Lastly, the proposed is computationally more  scalable because typical multi-output Gaussian processes have maximum $O(n^3k^3)$ complexity.}

There are a \textcolor{black}{few} limitations of this work. One  disadvantage is the computational cost \textcolor{black}{even though it increases linearly with $O(nk^3)$}. The proposed model incorporates more data from all bands and involves  more parameters to model their dependence. Thus, the resulting computational cost is inevitably expensive. For example, the CPU time taken for running the five-band simulation in Section~\ref{sec5.1} is about 26 hours, that for analyzing the SDSS five-band data in Section~\ref{sec5.2} is about 100 hours, and that for fitting the time delay model on the Q0957+561 data in Section~\ref{sec5.3} is about 9 hours on a personal laptop. \textcolor{black}{Streamlining} the code \textcolor{black}{implementation} is necessary to make it computationally less burdensome for fitting bigger data of large-scale surveys.  Another limitation is that the code for fitting the proposed model is for now  available only in R \citep{r2018}, although  astronomers are more familiar with Python. Thus, we plan to collaborate with bilingual astronomers or statisticians either to translate the R code to Python code or to develop a wrapper. \textcolor{black}{This  may also help reduce the computational cost.}

\textcolor{black}{In addition, it is unclear whether the proposed multivariate damped random walk is vulnerable to well-known limitations of univariate damped random walk. For example,}  \cite{mushotzky2011kepler} and \cite{zu2013is} report empirical evidence  that the damped random walk process fails to describe the optical variability of a quasar  on a very short timescale.  \cite{graham2014a} and \cite{kasliwal2015are} echo their arguments that not all types of AGN variabilities can be described by the damped random walk process. \textcolor{black}{\cite{kozlowski2016deg} warns of potential biases in associations between  model parameters and physical properties when true underlying processes are not damped random walk. Also, \cite{kozlowski2017limit} points out that a timescale estimate can be severely biased if a survey length is shorter than ten times the unknown true decorrelation timescale, which is typically the case in most surveys}. Looking into the limiting behavior of the process, \cite{tak2017bayesian}  demonstrate that it fails to fit AGN light curves if the  timescale of AGN variability is much smaller than the typical observation cadence in the data. \textcolor{black}{(See also \cite{kozlowski2017limit}  for simulation studies on this issue.)}

\textcolor{black}{It is unclear whether these limitations are equally applicable to a multivariate case.  This is mainly because various factors, such as the data quality in each band, can play a role in determining the accuracy of the resulting parameter estimation. In this work, for example, the dependence among multiple bands are modeled by their cross-correlation parameters. Thus, if the data quality is not good enough to estimate these correlations accurately, the proposed may not benefit from the idea of dependent modeling, i.e., sharing information across bands. An extensive simulation study that controls various data quality factors under  realistic settings \citep[e.g.,][]{kozlowski2017limit} is necessary for understanding the practical limitations of the proposed. We leave this as a future direction to explore.}


Several \textcolor{black}{more} opportunities to build upon the current work exist. (i) Since  multimodality is common in the time delay estimation, it is  promising to incorporate a multimodal Markov chain Monte Carlo sampler. For example, a repelling-attracting Metropolis algorithm has been successfully applied to a single-band time delay model  \citep{tak2018ram}. (ii) Next, the proposed model is stationary, meaning that it does not account for outlying observations. \cite{tak2019robust} demonstrate that the parameter estimation of a univariate damped random walk process can be severely biased in the presence of a few outliers. As an easy-to-implement solution, they introduce a computational trick that turns the Gaussian measurement errors to Student's $t$-errors, leading to more robust and accurate inferences. It will be a great improvement if this trick is incorporated into the proposed model fitting procedure. (iii) Also, improving the convergence rate of a Markov chain is important for enhancing computational efficiency. For a heteroscedastic model, it is theoretically shown that a specific data augmentation scheme can  expedite the convergence rate significantly \citep{tak2020dta}. Although this scheme has not been applied to heteroscedastic time series data, it is \textcolor{black}{well-worth investigating}. (iv) Finally, it is necessary to generalize the proposed multi-band model further. This is because the current model can describe only a specific type of variability defined by a damped random walk process, i.e., CARMA(1, 0). In a univarate case, however, a general-order CARMA($p, q$) model has been widely used due to  \textcolor{black}{limitations} of the damped random walk process \citep{kelly2014flexible, moreno2019stochastic}.  Thus, more flexible modeling on AGN variability will be possible by using the general-order multi-band CARMA($p, q$) model  \citep{marquardt2007mCARMA, schlemm2012multivariate}. We note that it is crucial to carefully design the state-space representation of the multi-band CARMA($p, q$)   to account for the heteroscedasticity of astronomical time series data. \textcolor{black}{We invite interested readers to explore these possibilities.}

\section{Concluding remarks}

In the era of astronomical big data with large-scale surveys, such as SDSS and LSST, it is important to \textcolor{black}{possess} various data analytic tools to handle the resulting multi-band data. There are possibly many existing tools that can be more powerful than the one presented in this work if a user is fully aware of their strong and weak points. For example, if fitting  smooth curves on multi-band time series data is of interest, various non-parametric curve-fitting methods such as kernel-smoothing, spline, wavelet, local  polynomial regression \citep{loader1999local, schafer2013} are available. However, these non-parameteric tools may not necessarily have interpretability that a parametric model can provide. A multivariate GP regression is a flexible way to model various types of variability,  e.g., incorporating a quasi-periodic aspect of the Doppler shift in a covariance function \citep{jones2017improving}. However, it is well known that a GP regression often involves a prohibitive  computational cost because it requires taking an inverse of a covariance matrix.  The tool we present in this work will be useful for modeling stochastic variability in irregularly-spaced astronomical time series data with heteroscedastic measurement errors, though it is not a panacea as described in Section~\ref{sec6}. We hope this work to initiate more active methodological research and discussion on multi-band data analyses in preparation for the upcoming era of LSST-driven time-domain astronomy.







\section*{Acknowledgements}
We appreciate  members of the International CHASC Astrostatistics Collaboration, especially Kaisey Mandel, David van Dyk, Vinay Kashyap, Xiao-Li Meng, and Aneta Siemiginowska, for  productive discussion on the first presentation of this work. Hyungsuk Tak appreciates Jogesh G.~Babu, Eric D.~Feigelson, and Eric \textcolor{black}{B.}~Ford for their insightful comments and  suggestions that have significantly improved the manuscript. Also, Hyungsuk Tak acknowledges computational support from the Institute for Computational and Data Sciences at Pennsylvania State University.  \textcolor{black}{Finally, the authors thank an anonymous referee for his or her thoughtful comments and feedback that enhanced the manuscript substantially.}

\software{\texttt{Rdrw} (Hu and Tak, 2020, version 1.0.0), 
\texttt{timedelay} (Tak, Mandel, van Dyk, Kashyap, Meng, Siemiginowska, 2015, version 1.0.10)
}


\bibliography{sample}{}

\begin{thebibliography}{}
\expandafter\ifx\csname natexlab\endcsname\relax\def\natexlab#1{#1}\fi
\providecommand{\url}[1]{\href{#1}{#1}}
\providecommand{\dodoi}[1]{doi:~\href{http://doi.org/#1}{\nolinkurl{#1}}}
\providecommand{\doeprint}[1]{\href{http://ascl.net/#1}{\nolinkurl{http://ascl.net/#1}}}
\providecommand{\doarXiv}[1]{\href{https://arxiv.org/abs/#1}{\nolinkurl{https://arxiv.org/abs/#1}}}

\bibitem[{{\'A}lvarez {et~al.}(2012){\'A}lvarez, Rosasco, \&
  Lawrence}]{alvarez2012kernels}
{\'A}lvarez, M., Rosasco, L., \& Lawrence, N. 2012, Kernels for Vector-Valued
  Functions: A Review, Foundations and Trends in Machine Learning Series (Now
  Publishers Incorporated)

\bibitem[{Andrae {et~al.}(2013)Andrae, Kim, \& Bailer-Jones}]{andrae2013}
Andrae, R., Kim, D.-W., \& Bailer-Jones, C. A.~L. 2013, A\&A, 554, A137,
  \dodoi{10.1051/0004-6361/201321335}

\bibitem[{Caruana(1997)}]{caruana1997}
Caruana, R. 1997, Machine Learning, 28, 41, \dodoi{10.1023/A:1007379606734}

\bibitem[{Czekala {et~al.}(2017)Czekala, Mandel, Andrews, Dittmann, Ghosh,
  Montet, \& Newton}]{czekala2017}
Czekala, I., Mandel, K.~S., Andrews, S.~M., {et~al.} 2017, The Astrophysical
  Journal, 840, 49, \dodoi{10.3847/1538-4357/aa6aab}

\bibitem[{Dobler {et~al.}(2015)Dobler, Fassnacht, Treu, Marshall, Liao,
  Hojjati, Linder, \& Rumbaugh}]{dobler2015}
Dobler, G., Fassnacht, C., Treu, T., {et~al.} 2015, The Astrophysical Journal,
  799, 168

\bibitem[{Durbin \& Koopman(2012)}]{durbin2012time}
Durbin, J., \& Koopman, S.~J. 2012, {Time Series Analysis by State Space
  Methods} (Oxford University Press)

\bibitem[{{Edelson} {et~al.}(2015){Edelson}, {Gelbord}, {Horne}, {McHardy},
  {Peterson}, {Ar{\'e}valo}, {Breeveld}, {De Rosa}, {Evans}, {Goad}, {Kriss},
  {Brandt}, {Gehrels}, {Grupe}, {Kennea}, {Kochanek}, {Nousek}, {Papadakis},
  {Siegel}, {Starkey}, {Uttley}, {Vaughan}, {Young}, {Barth}, {Bentz},
  {Brewer}, {Crenshaw}, {Dalla Bont{\`a}}, {De Lorenzo-C{\'a}ceres}, {Denney},
  {Dietrich}, {Ely}, {Fausnaugh}, {Grier}, {Hall}, {Kaastra}, {Kelly},
  {Korista}, {Lira}, {Mathur}, {Netzer}, {Pancoast}, {Pei}, {Pogge},
  {Schimoia}, {Treu}, {Vestergaard}, {Villforth}, {Yan}, \&
  {Zu}}]{edelson2015space}
{Edelson}, R., {Gelbord}, J.~M., {Horne}, K., {et~al.} 2015, The Astrophysical
  Journal, 806, 129, \dodoi{10.1088/0004-637X/806/1/129}

\bibitem[{Efron \& Morris(1975)}]{efron1975data}
Efron, B., \& Morris, C. 1975, Journal of the American Statistical Association,
  70, pp. 311.
\newblock \url{http://www.jstor.org/stable/2285814}

\bibitem[{Gardiner(2009)}]{gardiner2009}
Gardiner, C. 2009, {Stochastic Methods} (Berlin Heidelberg: Springer-Verlag)

\bibitem[{Gelfand {et~al.}(2004)Gelfand, Schmidt, Banerjee, \&
  Sirmans}]{gelfand2004}
Gelfand, A., Schmidt, A., Banerjee, S., \& Sirmans, C. 2004, TEST, 13, 263,
  \dodoi{10.1007/BF02595775}

\bibitem[{Graham {et~al.}(2014)Graham, Djorgovski, Drake, Mahabal, Chang,
  Stern, Donalek, \& Glikman}]{graham2014a}
Graham, M.~J., Djorgovski, S.~G., Drake, A.~J., {et~al.} 2014, Monthly Notices
  of the Royal Astronomical Society, 439, 703, \dodoi{10.1093/mnras/stt2499}

\bibitem[{Hojjati {et~al.}(2013)Hojjati, Kim, \& Linder}]{hojjati2013robust}
Hojjati, A., Kim, A.~G., \& Linder, E.~V. 2013, Physical Review D, 87, 123512

\bibitem[{Ivezi{\'{c}} {et~al.}(2019)Ivezi{\'{c}}, Kahn, Tyson, Abel, Acosta,
  Allsman, Alonso, AlSayyad, Anderson, Andrew, Angel, Angeli, Ansari,
  Antilogus, Araujo, Armstrong, Arndt, Astier, Aubourg, Auza, Axelrod, Bard,
  Barr, Barrau, Bartlett, Bauer, Bauman, Baumont, Bechtol, Bechtol, Becker,
  Becla, Beldica, Bellavia, Bianco, Biswas, Blanc, Blazek, Blandford, Bloom,
  Bogart, Bond, Booth, Borgland, Borne, Bosch, Boutigny, Brackett, Bradshaw,
  Brandt, Brown, Bullock, Burchat, Burke, Cagnoli, Calabrese, Callahan, Callen,
  Carlin, Carlson, Chandrasekharan, Charles-Emerson, Chesley, Cheu, Chiang,
  Chiang, Chirino, Chow, Ciardi, Claver, Cohen-Tanugi, Cockrum, Coles,
  Connolly, Cook, Cooray, Covey, Cribbs, Cui, Cutri, Daly, Daniel, Daruich,
  Daubard, Daues, Dawson, Delgado, Dellapenna, de~Peyster, de~Val-Borro, Digel,
  Doherty, Dubois, Dubois-Felsmann, Durech, Economou, Eifler, Eracleous,
  Emmons, Neto, Ferguson, Figueroa, Fisher-Levine, Focke, Foss, Frank, Freemon,
  Gangler, Gawiser, Geary, Gee, Geha, Gessner, Gibson, Gilmore, Glanzman,
  Glick, Goldina, Goldstein, Goodenow, Graham, Gressler, Gris, Guy, Guyonnet,
  Haller, Harris, Hascall, Haupt, Hernandez, Herrmann, Hileman, Hoblitt,
  Hodgson, Hogan, Howard, Huang, Huffer, Ingraham, Innes, Jacoby, Jain, Jammes,
  Jee, Jenness, Jernigan, Jevremovi{\'{c}}, Johns, Johnson, Johnson, Jones,
  Juramy-Gilles, Juri{\'{c}}, Kalirai, Kallivayalil, Kalmbach, Kantor, Karst,
  Kasliwal, Kelly, Kessler, Kinnison, Kirkby, Knox, Kotov, Krabbendam,
  Krughoff, Kub{\'{a}}nek, Kuczewski, Kulkarni, Ku, Kurita, Lage, Lambert,
  Lange, Langton, Guillou, Levine, Liang, Lim, Lintott, Long, Lopez, Lotz,
  Lupton, Lust, MacArthur, Mahabal, Mandelbaum, Markiewicz, Marsh, Marshall,
  Marshall, May, McKercher, McQueen, Meyers, Migliore, Miller, Mills, Miraval,
  Moeyens, Moolekamp, Monet, Moniez, Monkewitz, Montgomery, Morrison, Mueller,
  Muller, Arancibia, Neill, Newbry, Nief, Nomerotski, Nordby, O'Connor, Oliver,
  Olivier, Olsen, O'Mullane, Ortiz, Osier, Owen, Pain, Palecek, Parejko,
  Parsons, Pease, Peterson, Peterson, Petravick, Petrick, Petry, Pierfederici,
  Pietrowicz, Pike, Pinto, Plante, Plate, Plutchak, Price, Prouza, Radeka,
  Rajagopal, Rasmussen, Regnault, Reil, Reiss, Reuter, Ridgway, Riot, Ritz,
  Robinson, Roby, Roodman, Rosing, Roucelle, Rumore, Russo, Saha, Sassolas,
  Schalk, Schellart, Schindler, Schmidt, Schneider, Schneider, Schoening,
  Schumacher, Schwamb, Sebag, Selvy, Sembroski, Seppala, Serio, Serrano, Shaw,
  Shipsey, Sick, Silvestri, Slater, Smith, Smith, Sobhani, Soldahl,
  Storrie-Lombardi, Stover, Strauss, Street, Stubbs, Sullivan, Sweeney,
  Swinbank, Szalay, Takacs, Tether, Thaler, Thayer, Thomas, Thornton, Thukral,
  Tice, Trilling, Turri, Berg, Berk, Vetter, Virieux, Vucina, Wahl, Walkowicz,
  Walsh, Walter, Wang, Wang, Warner, Wiecha, Willman, Winters, Wittman, Wolff,
  Wood-Vasey, Wu, Xin, Yoachim, \& Zhan}]{ivezi2019lsst}
Ivezi{\'{c}}, {\v{Z}}., Kahn, S.~M., Tyson, J.~A., {et~al.} 2019, The
  Astrophysical Journal, 873, 111, \dodoi{10.3847/1538-4357/ab042c}

\bibitem[{Jones {et~al.}(2017)Jones, Stenning, Ford, Wolpert, Loredo, \&
  Dumusque}]{jones2017improving}
Jones, D.~E., Stenning, D.~C., Ford, E.~B., {et~al.} 2017, arXiv:1711.01318

\bibitem[{Journel \& Huijbregts(1978)}]{journel1978}
Journel, A.~G., \& Huijbregts, C.~J. 1978, {Mining Geostatistic} (London:
  Academic Press)

\bibitem[{Kalman(1960)}]{kalman1960}
Kalman, R.~E. 1960, Journal of Basic Engineering, 82, 35,
  \dodoi{10.1115/1.3662552}

\bibitem[{{Kasliwal} {et~al.}(2015){Kasliwal}, {Vogeley}, \&
  {Richards}}]{kasliwal2015are}
{Kasliwal}, V.~P., {Vogeley}, M.~S., \& {Richards}, G.~T. 2015, Monthly Notices
  of the Royal Astronomical Society, 451, 4328, \dodoi{10.1093/mnras/stv1230}

\bibitem[{Kelly {et~al.}(2009)Kelly, Bechtold, \&
  Siemiginowska}]{kelly2009variations}
Kelly, B.~C., Bechtold, J., \& Siemiginowska, A. 2009, The Astrophysical
  Journal, 698, 895

\bibitem[{Kelly {et~al.}(2014)Kelly, Becker, Sobolewska, Siemiginowska, \&
  Uttley}]{kelly2014flexible}
Kelly, B.~C., Becker, A.~C., Sobolewska, M., Siemiginowska, A., \& Uttley, P.
  2014, The Astrophysical Journal, 788, 33, \dodoi{10.1088/0004-637X/788/1/33}

\bibitem[{Kim {et~al.}(2012)Kim, Protopapas, Trichas, Rowan-Robinson, Khardon,
  Alcock, \& Byun}]{kim2012}
Kim, D.-W., Protopapas, P., Trichas, M., {et~al.} 2012, The Astrophysical
  Journal, 747, 107, \dodoi{10.1088/0004-637x/747/2/107}

\bibitem[{Koz{\l}owski(2016)}]{kozlowski2016deg}
Koz{\l}owski, S. 2016, Monthly Notices of the Royal Astronomical Society, 459,
  2787, \dodoi{10.1093/mnras/stw819}

\bibitem[{Koz{\l}owski {et~al.}(2010)Koz{\l}owski, Kochanek, Udalski,
  Soszy{\'n}ski, Szyma{\'n}ski, Kubiak, Pietrzy{\'n}ski, Szewczyk, Ulaczyk, \&
  Poleski}]{kozlowski2010quantifying}
Koz{\l}owski, S., Kochanek, C.~S., Udalski, A., {et~al.} 2010, The
  Astrophysical Journal, 708, 927

\bibitem[{Koz{\l}'owski(2017)}]{kozlowski2017limit}
Koz{\l}'owski, S. 2017, A\&A, 597, A128, \dodoi{10.1051/0004-6361/201629890}

\bibitem[{{Liao} {et~al.}(2015){Liao}, {Treu}, {Marshall}, {Fassnacht},
  {Rumbaugh}, {Dobler}, {Aghamousa}, {Bonvin}, {Courbin}, {Hojjati}, {Jackson},
  {Kashyap}, {Rathna Kumar}, {Linder}, {Mandel}, {Meng}, {Meylan}, {Moustakas},
  {Prabhu}, {Romero-Wolf}, {Shafieloo}, {Siemiginowska}, {Stalin}, {Tak},
  {Tewes}, \& {van Dyk}}]{kai2015}
{Liao}, K., {Treu}, T., {Marshall}, P., {et~al.} 2015, The Astrophysical
  Journal, 800, 11

\bibitem[{Loader(1999)}]{loader1999local}
Loader, C. 1999, {Local Regression and Likelihood} (New York: Springer-Verlag)

\bibitem[{MacLeod {et~al.}(2010)MacLeod, Ivezi{\'c}, Kochanek, Koz{\l}owski,
  Kelly, Bullock, Kimball, Sesar, Westman, Brooks, Gibson, Becker, \&
  de~Vries}]{macleod2010modeling}
MacLeod, C., Ivezi{\'c}, {\v{Z}}., Kochanek, C., {et~al.} 2010, The
  Astrophysical Journal, 721, 1014

\bibitem[{MacLeod {et~al.}(2012)MacLeod, Ivezi{\'{c}}, Sesar, de~Vries,
  Kochanek, Kelly, Becker, Lupton, Hall, Richards, Anderson, \&
  Schneider}]{macleod2012}
MacLeod, C.~L., Ivezi{\'{c}}, {\v{Z}}., Sesar, B., {et~al.} 2012, The
  Astrophysical Journal, 753, 106, \dodoi{10.1088/0004-637x/753/2/106}

\bibitem[{Marquardt \& Stelzer(2007)}]{marquardt2007mCARMA}
Marquardt, T., \& Stelzer, R. 2007, Stochastic Processes and their
  Applications, 117, 96, \dodoi{10.1016/j.spa.2006.05.014}

\bibitem[{Moreno {et~al.}(2019)Moreno, Vogeley, Richards, \&
  Yu}]{moreno2019stochastic}
Moreno, J., Vogeley, M.~S., Richards, G.~T., \& Yu, W. 2019, Publications of
  the Astronomical Society of the Pacific, 131, 063001,
  \dodoi{10.1088/1538-3873/ab1597}

\bibitem[{Morris(1983)}]{morris1983parametric}
Morris, C. 1983, Journal of the American Statistical Association, 78, pp. 47.
\newblock \url{http://www.jstor.org/stable/2287098}

\bibitem[{{Mushotzky} {et~al.}(2011){Mushotzky}, {Edelson}, {Baumgartner}, \&
  {Gand hi}}]{mushotzky2011kepler}
{Mushotzky}, R.~F., {Edelson}, R., {Baumgartner}, W., \& {Gand hi}, P. 2011,
  The Astrophysical Journal Letter, 743, L12,
  \dodoi{10.1088/2041-8205/743/1/L12}

\bibitem[{Nelder \& Mead(1965)}]{nelder1965}
Nelder, J.~A., \& Mead, R. 1965, The Computer Journal, 7, 308,
  \dodoi{10.1093/comjnl/7.4.308}

\bibitem[{{R Development Core Team}(2018)}]{r2018}
{R Development Core Team}. 2018, {R: A Language and Environment for Statistical
  Computing} (Vienna, Austria: R Foundation for Statistical Computing)

\bibitem[{Schafer \& Wasserman(2013)}]{schafer2013}
Schafer, C., \& Wasserman, L. 2013, Carnegie Mellon University

\bibitem[{Schlemm \& Stelzer(2012)}]{schlemm2012multivariate}
Schlemm, E., \& Stelzer, R. 2012, Bernoulli, 18, 46, \dodoi{10.3150/10-BEJ329}

\bibitem[{Shalyapin {et~al.}(2012)Shalyapin, Goicoechea, \&
  Gil-Merino}]{refId0}
Shalyapin, V.~N., Goicoechea, L.~J., \& Gil-Merino, R. 2012, Astronomy and
  Astrophysics, 540, A132, \dodoi{10.1051/0004-6361/201118316}

\bibitem[{Singh {et~al.}(2018)Singh, Ghosh, \& Adhikari}]{singh2018fast}
Singh, R., Ghosh, D., \& Adhikari, R. 2018, Physical Review E, 98, 012136,
  \dodoi{10.1103/PhysRevE.98.012136}

\bibitem[{Suyu {et~al.}(2017)Suyu, Bonvin, Courbin, Fassnacht, Rusu, Sluse,
  Treu, Wong, Auger, Ding, Hilbert, Marshall, Rumbaugh, Sonnenfeld, Tewes,
  Tihhonova, Agnello, Blandford, Chen, Collett, Koopmans, Liao, Meylan, \&
  Spiniello}]{suyu2017holicow}
Suyu, S.~H., Bonvin, V., Courbin, F., {et~al.} 2017, Monthly Notices of the
  Royal Astronomical Society, 468, 2590, \dodoi{10.1093/mnras/stx483}

\bibitem[{Tak {et~al.}(2019)Tak, Ellis, \& Ghosh}]{tak2019robust}
Tak, H., Ellis, J.~A., \& Ghosh, S.~K. 2019, Journal of Computational and
  Graphical Statistics, 28, 415, \dodoi{10.1080/10618600.2018.1537925}

\bibitem[{Tak {et~al.}(2018{\natexlab{a}})Tak, Ghosh, \& Ellis}]{tak2018how}
Tak, H., Ghosh, S.~K., \& Ellis, J.~A. 2018{\natexlab{a}}, Monthly Notices of
  the Royal Astronomical Society, 481, 277, \dodoi{10.1093/mnras/sty2326}

\bibitem[{Tak {et~al.}(2017{\natexlab{a}})Tak, Kelly, \& Morris}]{tak2017rgbp}
Tak, H., Kelly, J., \& Morris, C.~N. 2017{\natexlab{a}}, Journal of Statistical
  Software, 78, 1, \dodoi{10.18637/jss.v078.i05}

\bibitem[{Tak {et~al.}(2017{\natexlab{b}})Tak, Mandel, van Dyk, Kashyap, Meng,
  \& Siemiginowska}]{tak2017bayesian}
Tak, H., Mandel, K., van Dyk, D.~A., {et~al.} 2017{\natexlab{b}}, The Annals of
  Applied Statistics, 11, 1309, \dodoi{10.1214/17-AOAS1027}

\bibitem[{Tak {et~al.}(2018{\natexlab{b}})Tak, Meng, \& van Dyk}]{tak2018ram}
Tak, H., Meng, X.-L., \& van Dyk, D.~A. 2018{\natexlab{b}}, Journal of
  Computational and Graphical Statistics, 27, 479,
  \dodoi{10.1080/10618600.2017.1415911}

\bibitem[{Tak {et~al.}(2020)Tak, You, Ghosh, Su, \& Kelly}]{tak2020dta}
Tak, H., You, K., Ghosh, S.~K., Su, B., \& Kelly, J. 2020, Journal of
  Computational and Graphical Statistics, \dodoi{10.1080/10618600.2019.1704295}

\bibitem[{{Tewes} {et~al.}(2013){Tewes}, {Courbin}, \& {Meylan}}]{tewes2013a}
{Tewes}, M., {Courbin}, F., \& {Meylan}, G. 2013, Astronomy \& Astrophysics,
  553, A120

\bibitem[{Tierney(1994)}]{tierney1994markov}
Tierney, L. 1994, The Annals of Statistics, 22, 1701

\bibitem[{Treu \& Marshall(2016)}]{treu2016time}
Treu, T., \& Marshall, P.~J. 2016, The Astronomy and Astrophysics Review, 24,
  11

\bibitem[{Vatiwutipong \& Phewchean(2019)}]{vatiwutipong2019}
Vatiwutipong, P., \& Phewchean, N. 2019, Advances in Difference Equations,
  2918, 276, \dodoi{10.1186/s13662-019-2214-1}

\bibitem[{Zu {et~al.}(2016)Zu, Kochanek, Koz{\l}owski, \& Peterson}]{zu2016app}
Zu, Y., Kochanek, C.~S., Koz{\l}owski, S., \& Peterson, B.~M. 2016, The
  Astrophysical Journal, 819, 122, \dodoi{10.3847/0004-637x/819/2/122}

\bibitem[{{Zu} {et~al.}(2013){Zu}, {Kochanek}, {Koz{\l}owski}, \&
  {Udalski}}]{zu2013is}
{Zu}, Y., {Kochanek}, C.~S., {Koz{\l}owski}, s., \& {Udalski}, A. 2013, The
  Astrophysical Journal, 765, 106, \dodoi{10.1088/0004-637X/765/2/106}

\bibitem[{Zu {et~al.}(2011)Zu, Kochanek, \& Peterson}]{zu2011an}
Zu, Y., Kochanek, C.~S., \& Peterson, B.~M. 2011, The Astrophysical Journal,
  735, 80, \dodoi{10.1088/0004-637x/735/2/80}

\end{thebibliography}
\bibliographystyle{aasjournal}



\end{document}